\documentclass[pre,superscriptaddress,floatfix,notitlepage,nofootinbib]{revtex4-2}

\usepackage[utf8]{inputenc}
\usepackage{amsmath}
\usepackage{amssymb, bm}
\usepackage{stmaryrd}
\usepackage{lipsum}
\usepackage{bm}
\usepackage{graphicx}
\usepackage{graphics}
\usepackage[dvipdf]{epsfig}
\usepackage{subcaption}
\usepackage{float}
\usepackage{wrapfig}
\usepackage{url}
\usepackage{epstopdf} 
\usepackage{xcolor}
\usepackage{tensor} 
\usepackage{setspace}

\usepackage[format=plain,labelfont={bf,small},textfont=small,justification=raggedright,singlelinecheck=false]{caption}

\newcommand{\bd}{{\mathbf d}}
\newcommand{\bD}{{\mathbf D}}

\newcommand{\bQ}{{\mathbf{Q}}}

\newcommand{\com}[1]{{\color{black} #1}}
\newcommand{\comb}[1]{{\color{black} #1}}

\begin{document}

\title{A ruled narrow elastic strip model with corrected energy}

\author{E. Vitral}
\email{vitralfr@rose-hulman.edu}
\affiliation{Department of Mechanical Engineering, Rose-Hulman Institute of Technology, Terre Haute, IN 47803,  U.S.A.}
\author{J. A. Hanna}
\affiliation{Department of Mechanical Engineering, University of Nevada,
    1664  N.  Virginia  St.,  Reno,  NV  89557-0312,  U.S.A.}
\author{L. Koens}
\affiliation{School of Mathematical Sciences, Adelaide University, South Australia 5005 Australia}

\begin{abstract}
We present a new one-dimensional model for elastic strips based on a nondevelopable ruled surface. An auxiliary field regularizes the Sadowsky narrow-strip model to allow nonzero twist with vanishing curvature. 
The energy exhibits the scalings of the corrected functional derived by Freddi and co-workers, and for a certain choice of parameter, convexifies the Sadowsky energy without patching. We present the kinematics and energetics of the model, and employ a variational approach featuring a rotation tensor to derive equilibrium equations.  We perform a regular perturbation expansion to study the model behavior close to inflection points. When the energy is convex, curvature and moment are continuous at inflection points, while the auxiliary function suffers a jump, leading to a discontinuity in the ruled embedding for any finite width. 
\end{abstract}
\date{\today}


\maketitle


\section{Introduction}

Thin strips, also known as ribbons, occupy an interesting and challenging middle ground between rods and plates. 
One-dimensional strip models presume that both width-to-thickness $w/h \gg 1$ and length-to-width $L/w \gg 1$ ratios of the body are large.
Among them, we find the Sadowsky model~\cite{sadowsky1930elementarer,hinz2015translation} based on a narrow-width limit of an inextensible surface. 
Another is the Wunderlich model~\cite{wunderlich1962abwickelbares,todres2015translation} for finite-width strips, which employs a geometric construction based on developable ruled surfaces. This model reduces to Sadowsky in the limit of vanishing $w$. Further details about Wunderlich and its equilibrium solutions can be found in ~\cite{starostin2007shape,starostin2009force, starostin2015equilibrium}.
The surface inextensibility constraint limits the ribbon to isometric deformations, precluding internal wrinkles or creases~\cite{chopin2013helicoids,chopin2015roadmap,chopin2019extreme}. 
The relative simplicity of these models has attracted researchers interested in generalizing their application to more complex problems, such as incompatible ribbons~\cite{efrati2015non,grossman2016elasticity,hall2023building,gomez2023twisting} which do not possess a stress-free configuration, and diverse approaches towards numerical simulation of ribbons~\cite{moore2019computation,charrondiere2020numerical,neukirch2021convenient,huang2022discrete,charrondiere2024merci}. 

However, these classical models present serious issues that limit their range of application. 
The culprit is a ``twist over curvature'' term that forces twist to vanish whenever curvature vanishes, or curvature to jump between finite values, in order to avoid a singular energy. 
This causes problems in simulating any deformation that might introduce inflection points, and is an obstacle to connecting with the behavior of a simple anisotropic rod, which can display a straight, twisted centerline. 
The resulting predictions appear to be physically unrealistic at times \cite{yu2019bifurcations}, although we note that the  question of which models most reliably reproduce experiments and higher-dimensional simulations--- even in the qualitative sense of number of inflection points--- is complex and still very much an open area of research \cite{Kumar20, Kumar21}. 
The origin of this problematic term is the constraint of inextensibility of the surface, requiring a Gaussian curvature $K = 0$. 
The resulting energy is also non-convex.  If the energy is regularized to relax such constraints, then the strip may decompose into fine scale oscillations (``microstructure'') between a pair of solutions with small centerline curvature \cite{freddi2016corrected,paroni2019macroscopic,audoly2021one}. 

Mathematical results from Freddi and co-workers indicate that the narrow-width limit of a squared mean curvature energy for an inextensible sheet should actually scale differently depending on the amount of curvature, with a Sadowsky-like energy when curvature dominates over twist, and a Kirchhoff rod-like quadratic-twist energy when twist dominates \cite{freddi2016corrected,freddi2016variational}.  
Both Freddi and co-workers, and Audoly and Neukirch~\cite{audoly2021one}, convexify the Sadowsky energy in this manner by patching with a different form near inflection points.  

Recognizing that the Sadowsky model is unsuitable for the study of highly twisted ribbons, several groups have sought to relax the inextensibility constraint and thereby regularize inflection point singularities. One strategy is to add a small constant term meant to reflect stretching of the surface away from the centerline, which itself remains inextensible \cite{ghafouri2005helicoid,sano2019twist,audoly2021one,neukirch2021convenient}. 
However, these examples are either inconsistent with the Freddi scalings, or obtain them only by patching, rather than directly from a single energy bridging a high-curvature Sadowsky ribbon and a high-twist Kirchhoff rod. 
Other problems such as discontinuities in fields or geometric incompatibilities may also arise; further details are discussed in Section \ref{comparison}. 

Another option, which bears more of a resemblance to our approach, is Kumar and Rangarajan's introduction of an auxiliary field that also reflects some type of violation of surface inextensibility, for example through transverse curvature \cite{kumar2025centerline}.  Their model also allows for centerline stretching.  
These authors do not provide the corresponding equilibrium equations, and it is not immediately clear from their paper whether the resulting solutions connect with the Freddi energy. 

The ribbon model presented in this paper involves an auxiliary field that relaxes the developable Sadowsky model\com{, with its vanishing Gaussian curvature,} into a more general ruled surface. The surface stretches but the centerline remains inextensible. 
Equilibrium equations are derived concisely using a variational approach inherited from the shell literature \cite{wisniewski1998shell}. 
\comb{An appropriate parameter choice results in a convexified Sadowsky energy that reproduces the Freddi energy scalings in regions of high and low curvature without patching, although with different prefactors.}
The model admits a straight centerline solution with twist in the form of a helicoid. 
\com{However, while curvature and moment are continuous through inflection points under this parameter choice, the auxiliary function and any finite-width embedding suffer jumps.}
\comb{Another parameter choice matches the Freddi prefactors, but is nonconvex and suffers jumps in moment.}
 We derive an asymptotic bilinear form for the auxiliary function, of potential interest for numerical implementations. 
 
The paper is organized as follows. 
Section \ref{kinematics} presents the model kinematics and energy.
Euler-Lagrange equations of equilibrium using a rotation tensor are derived in Section \ref{EL}, with the general method also included in Appendix \ref{variational}.  
The remaining sections examine the behavior of the auxiliary field (Section \ref{regfield}), some limiting forms of the model (Section \ref{limits}), the scaling and landscape of the energy (Section \ref{energetics}), and a perturbation expansion of the continuous curvature/moment solution around inflection points (Section \ref{perturbation}).

\section{Strip kinematics and energy}\label{kinematics}

\subsection{Kinematics}

We begin by regularizing the embedding of a rectifying developable into a particular form of nondevelopable ruled surface. 
 Let $\mathbf{y}(s,\nu)$ denote a point on the present configuration of a ribbon, parametrized by the arc length $s$ along its centerline $\mathbf{x}(s)$, and by a coordinate $\nu$ along a generator: 
\begin{align}
  \mathbf{y}(s,\nu) &= \mathbf{x}(s)+\nu\big[\mathbf{d}_1(s)+f(s)\mathbf{d}_3(s)\big]\,,\label{embedding}
  \\[3mm]
  f(s) &= \frac{\tau(s)}{\kappa(s)+\epsilon(s)}\,.  \label{eq:f}
\end{align}
Here, the $\{\mathbf{d}_i\}$ form a right-handed orthonormal basis on the centerline, with $\mathbf{d}_2$ normal to the surface and $\mathbf{d}_3 = \mathbf{x}'$ tangent to the inextensible centerline (the prime denoting differentiation with respect to arc length $s$). Without the independent regularizing field $\epsilon(s)$, we would have the classical Sadowsky kinematics. The fields $\kappa (s)$ and $\tau (s)$ describe the rotation of the frame, akin to the kinematics of a perfectly anisotropic rod with one forbidden curvature direction:
\begin{align}
  \label{eq:kin}
  \begin{split}
  \mathbf{d}_1' &= \tau\mathbf{d}_2 \,,
  \\[3mm]
  \mathbf{d}_2' &= \kappa\mathbf{d}_3-\tau\mathbf{d}_1 \,,
  \\[3mm]
  \mathbf{d}_3' &= -\kappa\mathbf{d}_2 \,,
  \end{split}
\end{align}
that is, $\kappa_2 \equiv  -\mathbf{d}_1'\cdot \mathbf{d}_3 = \mathbf{d}_3'\cdot \mathbf{d}_1 = 0$. 
There is a direction, material on the centerline, in which the centerline does not turn. The Frenet frame coincides with the material frame on the centerline only. \comb{ The field $\tau = \mathbf{d}_1'\cdot\mathbf{d}_2 = -\mathbf{d}_2'\cdot\mathbf{d}_1$ is the material twist of the ribbon on the centerline.} 
 The tangent plane on the centerline lies in the rectifying plane. However, the entire surface does not lie in this plane; the normals rotate along the generators, and the surface curls away.

For the purposes of deriving equilibrium equations using a convenient variational approach, we may also describe the kinematics in terms of a rotation tensor $\mathbf{Q}\in \textrm{SO}(3)$. 
This rotation takes a reference orthonormal basis for the strip centerline $\{\mathbf{D}_i\}$ to $\{\mathbf{d}_i\}$ by
\begin{equation}
  \label{eq:q}
  \mathbf{d}_i = \mathbf{Q}\cdot\mathbf{D}_i\,.
\end{equation}
Differentiation of \eqref{eq:q} leads to
\begin{equation}
    \mathbf{d}_i' = \mathbf{Q}'\cdot\mathbf{D}_i = \mathbf{Q}'\cdot\mathbf{Q}^\top\cdot\mathbf{d}_i = \mathrm{ax}\left(\mathbf{Q}'\cdot\mathbf{Q}^\top\right) \times \mathbf{d}_i \, ,
\end{equation}
where $\mathbf{Q}'\cdot\mathbf{Q}^\top$ is a skew-symmetric tensor whose axial vector is
\begin{equation}
\mathrm{ax}\left(\mathbf{Q}'\cdot\mathbf{Q}^\top\right) = -(\kappa \mathbf{d}_1 + \tau \mathbf{d}_3).
    \label{eq:curv}
\end{equation}

\subsection{Narrow strip energy}

We consider an energy in the Sadowsky limit of vanishing width by deriving quantities for the surface \eqref{embedding} and using their values on the centerline ($\nu = 0$).
 Based on the derivatives 
\begin{align}
  \partial_s\mathbf{y}
  &= \mathbf{d}_3\,,\quad
    \partial_\nu\mathbf{y} = \mathbf{d}_1+f\mathbf{d}_3\,,
\end{align}
we can calculate the following covariant $a_{\alpha\beta}$ and contravariant $a^{\alpha\beta}$ metric components and determinant $a$
\begin{align}
    a_{ss} &= 1\,,\quad a_{s\nu} = f\,,\quad a_{\nu\nu} = 1+f^2\,,\quad a= 1\,,
    \\[2mm]
    a^{ss} &= 1+f^2\,,\quad a^{s\nu} = -f\,,\quad a^{\nu\nu} = 1\,,
\end{align}
along with the curvature components $b_{\alpha\beta}$, and mean $H$ and Gaussian $K$ curvature invariants
\begin{align}
    b_{ss} &= -\kappa\,,\quad b_{s\nu} = \epsilon f\,,\quad b_{\nu\nu} = 0\,,
    \\[2mm]
    \label{eq:mc}
    H &= -\epsilon f^2-\tfrac{1}{2}\kappa(1+f^2)\,,
    \\[2mm]
    \label{eq:gc}
    K &= -\epsilon^2f^2\,.
\end{align}



In the narrow-width limit, we approximate all terms with their centerline values and write a phenomenological one-dimensional energy density quadratic in mean curvature and $\epsilon$, the latter serving to represent Gaussian contributions to both stretching and bending.  We assign Lagrange multipliers $\mathbf{N}$ and $M_2$ to respectively enforce the constraints of (one-dimensional) inextensibility $\mathbf{x}' = \mathbf{d}_3$ and vanishing forbidden curvature $\kappa_2 = 0$,  
\begin{align}
  \label{eq:energy}
  \mathcal{E} &= \int ds \, e = \int ds \bigg[
  \frac{b_{\epsilon}}{2}\epsilon^2+\frac{b_H}{2}H^2+M_2\kappa_2
  +\mathbf{N}\cdot(\mathbf{x}'-\mathbf{d}_3)
  \bigg] \,,
\end{align}
where $e$ is the energy density and $b_{\epsilon}$ and $b_H$ are constant moduli. 

Recalling the form of $f$ from \eqref{eq:f}, we note that when $\epsilon \rightarrow 0$, this bending energy is the Sadowsky energy, scaling like $\left(\kappa^2+\tau^2\right)^2/\kappa^2$.  When $\kappa \rightarrow 0$, the bending energy scales like $\tau^4/\epsilon^2$, so should we find that $\epsilon \sim \tau$ in this limit, both the bending and stretching energies will scale like $\tau^2$, consistent with \cite{freddi2016corrected}.  Note that $\epsilon$ need not be small, and when $\kappa \rightarrow 0$ the Gaussian curvature simply scales as $\tau^2$ regardless of the behavior of $\epsilon$.

In the next section, we will connect the Lagrange multipliers to the moment $\mathbf{M}$ and force $\mathbf{N}$ vectors
\begin{equation}
  \mathbf{M} = M_1\mathbf{d}_1+M_2\mathbf{d}_2+M_3\mathbf{d}_3
  \,,\quad \mathbf{N} = N_1\mathbf{d}_1+N_2\mathbf{d}_2+N_3\mathbf{d}_3\,.
\end{equation}

\section{Euler-Lagrange equations}\label{EL}
\label{sec:euler}

We now derive the Euler-Lagrange equations of equilibrium for this regularized ribbon model.
Instead of following the usual variational derivations for nonlinear elastic rods~\cite{steigmann1993variational,dias2015wunderlich,paroni2019macroscopic}, we employ a convenient approach from Wi{\'s}niewski~\cite{wisniewski1998shell}, originally developed for shells, that employs variations of a rotation tensor. A general example of this approach for an elastic rod is detailed in Appendix~\ref{variational}.

The variation of the rotation $\mathbf{Q}$ is characterized in terms of a pseudovector $\delta\boldsymbol\theta$, defined through the relation $\delta\mathbf{Q}\cdot\mathbf{Q}^\top\cdot () \equiv \delta\boldsymbol\theta\times ()$. From~\eqref{eq:q}, we have both $\mathbf{d}_i = \mathbf{Q}\cdot\mathbf{D}_i$ (so that $\delta \mathbf{d}_i = \delta \mathbf{Q}\cdot \mathbf{D}_i$) and $\mathbf{D}_i = \mathbf{Q}^\top\cdot\mathbf{d}_i$. For example,
\begin{equation}
\mathbf{N}\cdot\delta\mathbf{d}_3 = \mathbf{N}\cdot\delta\mathbf{Q}\cdot\mathbf{Q}^\top\cdot\mathbf{d}_3=\mathbf{N}\cdot\delta\boldsymbol\theta\times\mathbf{d}_3 = \mathbf{d}_3\times\mathbf{N}\cdot\delta\boldsymbol\theta\,.
\label{eq:nd3}
\end{equation}
Relating with~\eqref{eq:kin}, we find that variations of the frame evolution fields can be expressed as
\begin{align}
  \label{eq:varcurv}
  \begin{split}
    \delta\tau &= (\delta\boldsymbol\theta\times\mathbf{d}_1)'\cdot\mathbf{d}_2
    +\mathbf{d}_1'\cdot(\delta\boldsymbol\theta\times\mathbf{d}_2)\,,
    \\[2mm]
    \delta\kappa &= (\delta\boldsymbol\theta\times\mathbf{d}_2)'\cdot\mathbf{d}_3
    +\mathbf{d}_2'\cdot(\delta\boldsymbol\theta\times\mathbf{d}_3)\,,
    \\[2mm]
    \delta\kappa_2 &= (\delta\boldsymbol\theta\times\mathbf{d}_3)'\cdot\mathbf{d}_1
    +\mathbf{d}_3'\cdot(\delta\boldsymbol\theta\times\mathbf{d}_1)\,.
  \end{split}
\end{align}
The first variation of the ribbon energy~\eqref{eq:energy} is
\begin{align}
  \label{eq:1v}
  \begin{split}
  \delta\mathcal{E} = \int ds\bigg\{
  & (-b_HHf^2+ b_\epsilon \epsilon)\delta\epsilon
    -b_H\frac{H}{2}(1+f^2)\delta\kappa
  -b_HH(2\epsilon f+\kappa f)\delta f+ M_2\delta\kappa_2
  + \mathbf{N}\cdot\delta\mathbf{x}'
    -\mathbf{d}_3\times\mathbf{N}\cdot\delta\boldsymbol\theta
    \bigg\}\,.
  \end{split}
\end{align}
We can make use of~\eqref{eq:f} to rewrite $\delta f$ as
\begin{equation}
  \delta f = \frac{1}{\kappa+\epsilon}\delta\tau
  -\frac{\tau}{(\kappa+\epsilon)^2}(\delta\kappa+\delta\epsilon)
  = \frac{\delta\tau - f (\delta\kappa+\delta\epsilon)}{\kappa+\epsilon} \, .
\end{equation}
Based on the previous expression, we cast~\eqref{eq:1v} in the form
\begin{align}
  \label{eq:1v2}
  \delta\mathcal{E}
  &= \int ds \bigg(
    \alpha \delta\epsilon+M_3\delta\tau+M_1\delta\kappa
    +M_2\delta\kappa_2
    +\mathbf{N}\cdot\delta\mathbf{x}'
    -\mathbf{d}_3\times\mathbf{N}\cdot\delta\boldsymbol\theta
    \bigg)\, ,
\end{align}
where the conjugate quantities to $\epsilon$, $\kappa$, and $\tau$ are
\begin{align}
  \label{eq:moment1}
  \begin{split}
    \alpha  &= b_\epsilon\epsilon+b_H H \,\frac{\tau^2\epsilon}{(\kappa+\epsilon)^3}
    \,,
    \\[2mm]
    M_1 &= b_H H \, \bigg[ \frac{1}{2}\bigg( \frac{\tau^2}{(\kappa+\epsilon)^2}-1\bigg)
    +\frac{\epsilon\tau^2}{(\kappa+\epsilon)^3}\bigg]
    \,,
    \\[2mm]
    M_3 &= -b_H H \,\frac{\tau(\kappa+2\epsilon)}{(\kappa+\epsilon)^2}    
    \,.
  \end{split}
\end{align}
Substituting the variations from~\eqref{eq:varcurv} into~\eqref{eq:1v2}, and accounting for the orthogonality between basis vectors, we arrive at the form
\begin{align}
  \delta\mathcal{E}
  & = \int ds \bigg[
    \alpha\delta\epsilon
    -(\mathbf{M}'+\mathbf{x}'\times\mathbf{N})\cdot\delta\boldsymbol\theta
    -\mathbf{N}'\cdot\delta\mathbf{x} \bigg]
    +(\mathbf{M}\cdot\delta\boldsymbol\theta+\mathbf{N}\cdot\delta\mathbf{x})\big|^L_0 \,.
\end{align}
By satisfying $\delta\mathcal{E} = 0$ for all possible variations, we obtain the Euler-Lagrange equations in the absence of external forces and moments: 
\begin{align}
  \label{eq:ep}
  &\alpha = b_\epsilon\epsilon-b_H\frac{\tau^2\epsilon}{(\kappa+\epsilon)^5} \bigg[\epsilon \tau^2+\frac{\kappa}{2}(\kappa^2+2\kappa\epsilon+\epsilon^2+\tau^2)\bigg]
  = b_\epsilon\epsilon-b_H \bigg[\epsilon f^2+\frac{\kappa}{2}(1+f^2)\bigg] \frac{f^2\epsilon}{\kappa+\epsilon}
  = 0\,,
  \\[2mm]
  \label{eq:blm}
  &\mathbf{N}' = \mathbf{0}\,,
  \\[2mm]
  \label{eq:bam}
  &\mathbf{M}'+\mathbf{x}'\times\mathbf{N} = \mathbf{0}\,.
\end{align}

\section{Regularizing field}\label{regfield}

We are interested in how inflection point singularities are regularized such that $\kappa$ vanishes and $\tau$ does not. To this end, we examine the algebraic relation \eqref{eq:ep} between the auxiliary field $\epsilon$ and the frame evolution fields $\kappa$ and $\tau$.

\subsection{Real solutions}

\begin{figure}[ht]
    \centering
    \begin{subfigure}{0.4\textwidth}    
    \includegraphics[width=1\linewidth]{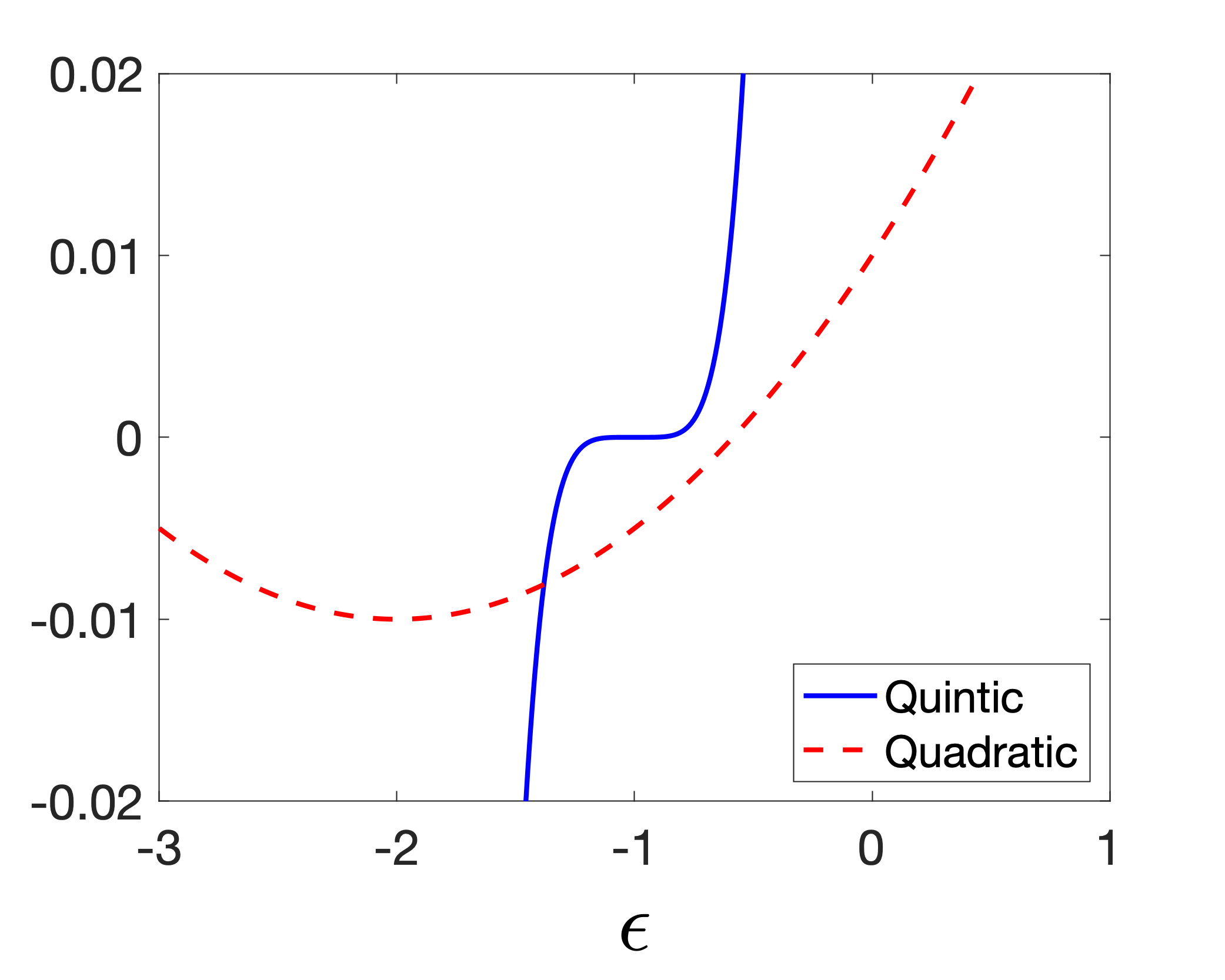} \quad
      \caption{$b_H/b_\epsilon = 0.01$}
    \end{subfigure}
    \begin{subfigure}{0.4\textwidth}    
    \includegraphics[width=1\linewidth]{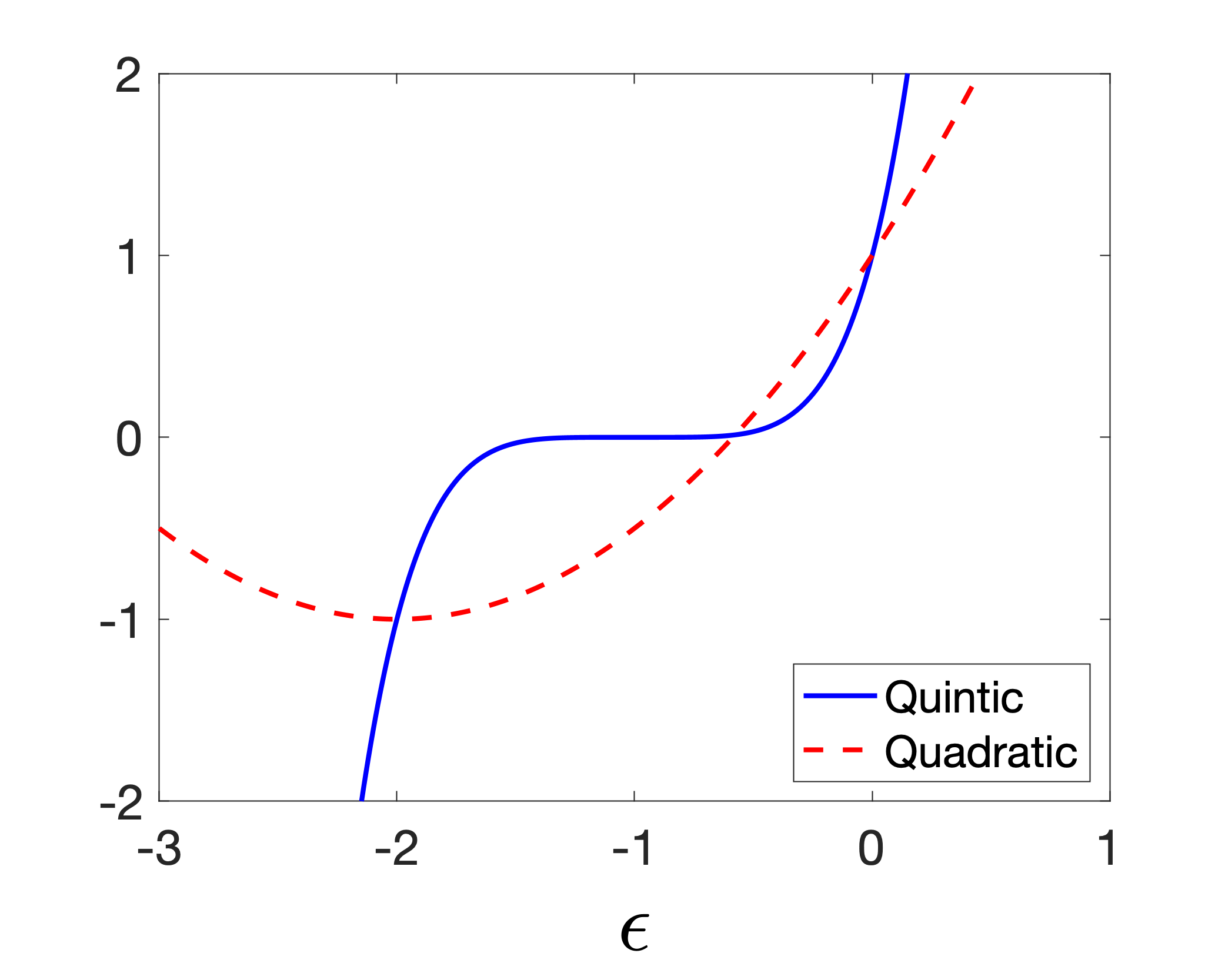} \quad
      \caption{$b_H/b_\epsilon = 1$}
    \end{subfigure}
    \caption{Quintic (LHS) and quadratic (RHS) functions of $\epsilon$ from different sides of Eq.~\eqref{eq:roots}, with $\kappa=1$ and $\tau=1$, showing one real root for $b_H/b_\epsilon = 0.01$ and three for $b_H/b_\epsilon=1$.}
    \label{fig:epsroot}    
\end{figure}

Six solutions for $\epsilon(\kappa,\tau)$ can be obtained from~\eqref{eq:ep}, including a trivial one, $\epsilon=0$. Assuming $\epsilon \neq 0$ and $\kappa \neq 0$, the equation can be rearranged as
\begin{equation}
  (\kappa+\epsilon)^5 = \frac{b_H}{b_\epsilon}\frac{\kappa\tau^2}{2}\bigg[\bigg(\epsilon+\frac{\kappa^2+\tau^2}{\kappa}\bigg)^2- \frac{\tau^2(\kappa^2+\tau^2)}{\kappa^2}\bigg]\,. \label{eq:roots}
\end{equation}
 The above equation shows that the roots of $\epsilon$ are the intersection of a simple quintic and a quadratic.  
 This equation for $\epsilon$ has one real root when $b_H/b_\epsilon$ is small, and three real roots otherwise. Figure~\ref{fig:epsroot} shows examples of each type.

 For fixed $\tau$, as $\kappa \to 0 ^{\pm}$, the turning point of the quadratic approaches $(\pm\infty, \mp\infty)$. This discontinuity corresponds to a discontinuity in the always-present real root as $\kappa$ goes through zero, reflective of a jump in $\epsilon$ around inflection points, which will be discussed in greater detail in Sec.~\ref{sec:asymptotic}. 
The transition between one and three real roots occurs when Eq.~\eqref{eq:roots} is satisfied along with its derivative with respect to $\epsilon$, meaning that the quintic and quadratic are tangent:
\begin{equation}
	5 (\kappa+\epsilon)^4 = \frac{b_H}{b_\epsilon} \kappa\tau^2\bigg(\epsilon+\frac{\kappa^2+\tau^2}{\kappa}\bigg)\,. 
\end{equation}
Rearranging these equations, we find that the transition occurs when
\begin{equation}
 (\kappa+\epsilon) \bigg(\epsilon+\frac{\kappa^2+\tau^2}{\kappa}\bigg)=  \frac{5}{2}\bigg[\bigg(\epsilon+\frac{\kappa^2+\tau^2}{\kappa}\bigg)^2- \frac{\tau^2(\kappa^2+\tau^2)}{\kappa^2}\bigg] \,,\quad \textrm{and}\quad \frac{b_H}{b_\epsilon} = \frac{5(\kappa+\epsilon)^4}{\kappa\tau^2(\epsilon+\frac{\kappa^2 +\tau^2}{\kappa})}\,.
\end{equation}
 The first equation provides the location of $\epsilon$ at which the transition occurs as a function of $\kappa$ and $\tau$, while the second equation provides the values of $b_H/b_\epsilon$ required for the transition to occur.

\subsection{Vanishing curvature solution}

In the limit of vanishing curvature $\kappa\rightarrow 0$, we find the following pair of solutions $\epsilon = \epsilon_0$ from \eqref{eq:ep}, 
\begin{equation}
  \label{eq:ep0}
  \epsilon_0 = \pm \bigg(\frac{b_H}{b_\epsilon}\bigg)^{1/4}\tau\,. 
\end{equation}
Therefore, in this limit, the Gaussian curvature~\eqref{eq:gc} is $K = -\tau^2$, as in a helicoid, and the energy density from~\eqref{eq:energy} scales quadratically with twist $e \sim \tau^2$. This is similar to the corrected Sadowsky energy of Freddi and co-workers ~\cite{freddi2016corrected}.

\subsection{Asymptotic solution} \label{sec:asymptotic}

Away from the jump at $\kappa = 0$, the regularizing field shows an abrupt transition (with discontinuous first derivative) between zero and a linear trend with $\kappa$ (this can be seen in Figs.~\ref{fig:energy1} and \ref{fig:energy2} in a later Section (Sec. \ref{energetics}) alongside a discussion of the energy landscape). 
When $\epsilon = 0$, the model reduces to the classical Sadowsky ribbon. We are interested in understanding how $\epsilon$ changes between zero in Sadowsky-like regions of dominant curvature and $\epsilon_0$ in regions of dominant twist.  For this analysis, it is helpful to derive an asymptotic form for $\epsilon$ in the limit of small $\kappa$. At order $\kappa$, \eqref{eq:roots} becomes
\begin{equation}
    2b_\epsilon(\epsilon^5+5\kappa\epsilon^4)-b_H\tau^2[2\epsilon\tau^2+\kappa(\epsilon^2+\tau^2)]  = O(\kappa^2)\,.
    \label{eq:epksmall}
\end{equation}
Note that we have already divided by $\epsilon$, accounting for the trivial solution. We now look for linear solutions close to~\eqref{eq:ep0}, of the form
\begin{equation}
    \epsilon_1 = \beta\kappa\pm\bigg(\frac{b_H}{b_\epsilon}\bigg)^{1/4}\tau\,.
    \label{eq:ep1}
\end{equation}
By substituting~\eqref{eq:ep1} into~\eqref{eq:epksmall} and dropping higher order terms, we obtain
\begin{eqnarray}
  2b_\epsilon\bigg[\bigg(\frac{b_H}{b_\epsilon}\bigg)^{5/4}\tau^4+5\frac{b_H}{b_\epsilon}\kappa\tau^4(\beta+1)\bigg]
  -b_H\tau^2\bigg[2\bigg(\frac{b_H}{b_\epsilon}\bigg)^{1/4}\tau^2+2\beta\kappa\tau^2+\kappa\tau^2\bigg(\bigg(\frac{b_H}{b_\epsilon}\bigg)^{1/2}+1\bigg)\bigg] = O(\kappa^2) \,.
\end{eqnarray}
After cancellations, we find
\begin{equation}
    \beta = \frac{1}{8}\bigg[\bigg(\frac{b_H}{b_\epsilon}\bigg)^{1/2}\hspace{-1mm}-9\bigg]\,. 
\end{equation}
This result suggests that an approximate solution which bridges the Sadowsky energy in regions where $\kappa$ dominates over $\tau$, to a regularized form in regions where $\tau$ dominates over $\kappa$, is
\begin{equation}
\label{eq:epasym}
  \epsilon(\kappa,\tau) =
  \begin{cases}
    \dfrac{1}{8}\bigg[\bigg(\dfrac{b_H}{b_\epsilon}\bigg)^{1/2}\hspace{-1mm}-9\bigg]\kappa -\bigg(\dfrac{b_H}{b_\epsilon}\bigg)^{1/4}\tau &\text{for}\quad (\kappa <0\,,\, \tau > 0\,,\, \epsilon \leq 0) \quad\text{or}\quad (\kappa>0\,,\,\tau<0\,,\,\epsilon\geq 0)\;,\\[5mm]
    \dfrac{1}{8}\bigg[\bigg(\dfrac{b_H}{b_\epsilon}\bigg)^{1/2}\hspace{-1mm}-9\bigg]\kappa +\bigg(\dfrac{b_H}{b_\epsilon}\bigg)^{1/4}\tau &\text{for}\quad (\kappa <0\,,\, \tau < 0\,,\, \epsilon \leq 0) \quad\text{or}\quad (\kappa>0\,,\,\tau>0\,,\,\epsilon\geq 0)\,, \\[5mm]
    0 & \text{otherwise.}
  \end{cases}       
\end{equation}
 For a fixed $\tau$,
this equation tells us that the $\epsilon$ field is zero for large curvature, and linear otherwise. The linear coefficient of the equation is the same for both $\kappa < 0$ to $\kappa > 0$, so there must be a discontinuity in $\epsilon$ at $\kappa = 0$. This discontinuity is a jump of $\llbracket \epsilon \rrbracket = 2(b_H/b_\epsilon)^{1/4}\tau$ at inflection points $\kappa = 0$, which is the difference between the two $\epsilon_0$ solutions in \eqref{eq:ep0}, \com{and is present for any nonzero choice of $b_H$ and $b_\epsilon$. However, the expression \eqref{eq:gc} representing the centerline Gaussian curvature remains continuous, as it is proportional to $\epsilon^2$ and, furthermore, $K=-\tau^2$ at inflection points ($\kappa=0$).}

\subsection{Embedding jump}

The jump in $\epsilon$ at inflection points implies a jump in the embedding away from the centerline, which is given by
\begin{equation}
  \label{eq:jump}
    \llbracket\mathbf{y}\rrbracket = 2\nu\bigg(\frac{b_\epsilon}{b_{H}}\bigg)^{1/4}\mathbf{d}_3\,.
\end{equation}
While this vanishes in the Sadowsky limit of vanishing width, for real ribbons of small finite width, a jump in $\mathbf{y}$ and in the orientation of the generators will appear at inflection points. This will be discussed further in Sec.~\ref{perturbation}.

\subsection{Note on numerical implementation}

The bilinear expressions \eqref{eq:epasym} offer a good approximation for the behavior of the regularizing field, so can be conveniently used to calculate $\epsilon$ and its derivative with respect to $s$ in a numerical implementation of the model. 
For example, we can determine the desirable root of~\eqref{eq:ep} based on which one is the closest to~\eqref{eq:epasym}. In addition, \eqref{eq:epasym} and its derivative can be substituted in the governing equations~\eqref{eq:force} and \eqref{eq:moment3} to eliminate the dependence on the regularizing field and find an approximate solution for the relevant quantities. In such an approach, $\epsilon$ would be determined pointwise through $(\kappa,\tau)$, and $\epsilon'$ just modifies the coefficients of $\kappa'$ and $\tau'$ in the ODEs obtained from the balance equations.

\section{Model limits}\label{limits}

Based on the derived solutions for $\epsilon$, we are ready to analyze the balance of linear momentum~\eqref{eq:blm} and balance of angular momentum~\eqref{eq:bam} for different limits of the model's variables.
In component form, the balance of linear momentum~\eqref{eq:blm} is
\begin{align}
  \begin{split}
    N_1'-N_2\tau &= 0\,,
    \\[2mm]
    N_2' + N_1\tau -N_3\kappa &= 0\,,
    \\[2mm]
    N_3' + N_2\kappa &= 0 \,,
    \end{split}
    \label{eq:force2}
\end{align}
and the balance of angular momentum~\eqref{eq:bam} is 
\begin{align}
  \begin{split}
    M_1'-M_2\tau-N_2 &= 0\,,
    \\[2mm]
    M_2'+M_1\tau -M_3\kappa + N_1 &= 0\,,
    \\[2mm]
    M_3' + M_2\kappa  &= 0\,.
  \end{split}
  \label{eq:moment2}
\end{align}

Linear constitutive laws for a perfectly anisotropic rod are of the type $M_1^{rod} = \alpha_1\kappa$ and $M_3^{rod} = \alpha_3 \tau$, where the $\alpha_i$ are constants, with $M_2^{rod}$ a multiplier determined by the constraint of vanishing $\kappa_2$.  
We are interested in understanding how the present model bridges the rod equations to the Sadowsky strip model. For this, we evaluate how $M_1$ and $M_3$ behave in three distinct limits. This analysis is based on the expressions in~\eqref{eq:moment1}, which upon inserting for mean curvature are
\begin{align}
  \begin{split}
    M_1 &= b_H\bigg[-\frac{\epsilon\tau^2}{(\kappa+\epsilon)^2}
    -\frac{\kappa}{2}\bigg(1+\frac{\tau^2}{(\kappa+\epsilon)^2}\bigg) \bigg]
    \bigg[\frac{1}{2}\bigg(\frac{\tau^2}{(\kappa+\epsilon)^2}-1\bigg)
    +\frac{\epsilon\tau^2}{(\kappa+\epsilon)^3}\bigg]
    \,,
    \\[2mm]
    M_3 &= -b_H \bigg[-\frac{\epsilon\tau^2}{(\kappa+\epsilon)^2}
    -\frac{\kappa}{2}\bigg(1+\frac{\tau^2}{(\kappa+\epsilon)^2}\bigg) \bigg]
    \frac{\tau(\kappa+2\epsilon)}{(\kappa+\epsilon)^2}
    \,.
  \end{split}
  \label{eq:moment3}
\end{align}


\subsection{Limit $\epsilon\rightarrow 0$}
In the limit $\epsilon \rightarrow 0$, we find
\begin{align}
  \begin{split}
    M_1 &= \frac{b_H}{4}\kappa \bigg(1-\frac{\tau^4}{\kappa^4}\bigg)
    \,,
    \\[2mm]
    M_3 &= \frac{b_H}{2}\tau\bigg(1+\frac{\tau^2}{\kappa^2}\bigg)
    \,.
  \end{split}
\end{align}
In this limit, we recover the Sadowsky model for a strip. The main difference between this model and the anisotropic rod model is the presence of a $\kappa f^2$ contribution to the mean curvature $H$, which manifests in $M_1$ and $M_3$ as a ``$\tau$ over $\kappa$'' term. From~\eqref{eq:epasym}, we see that $\epsilon = 0$ will be a solution approximately when 
\begin{equation}
    |\tau| \lesssim    \frac{1}{8}\bigg[9\bigg(\frac{b_\epsilon}{b_H}\bigg)^{1/4}-\bigg(\frac{b_H}{b_\epsilon}\bigg)^{1/4}\bigg]|\kappa|\,. 
    \label{eq:zero}
\end{equation}
When $\tau \ll \kappa$ (small $f$), this model reduces to the anisotropic rod equations, for which $M_1 \sim \kappa$ and $M_3 \sim \tau$. 


\subsection{Limit $\tau\rightarrow 0$}

In the limit $\tau \rightarrow 0$, we find
\begin{align}
  \begin{split}
    M_1 &= \frac{b_H}{4}\kappa
    \,,
    \\[2mm]
    M_3 &= 0
    \,,
  \end{split}
\end{align}
which trivially behaves like a rod, since $f = 0$ and $H = -\kappa/2$.


\subsection{Limit $\kappa\rightarrow 0$}

\begin{figure}[ht]
    \centering
    \includegraphics[width=0.33\linewidth]{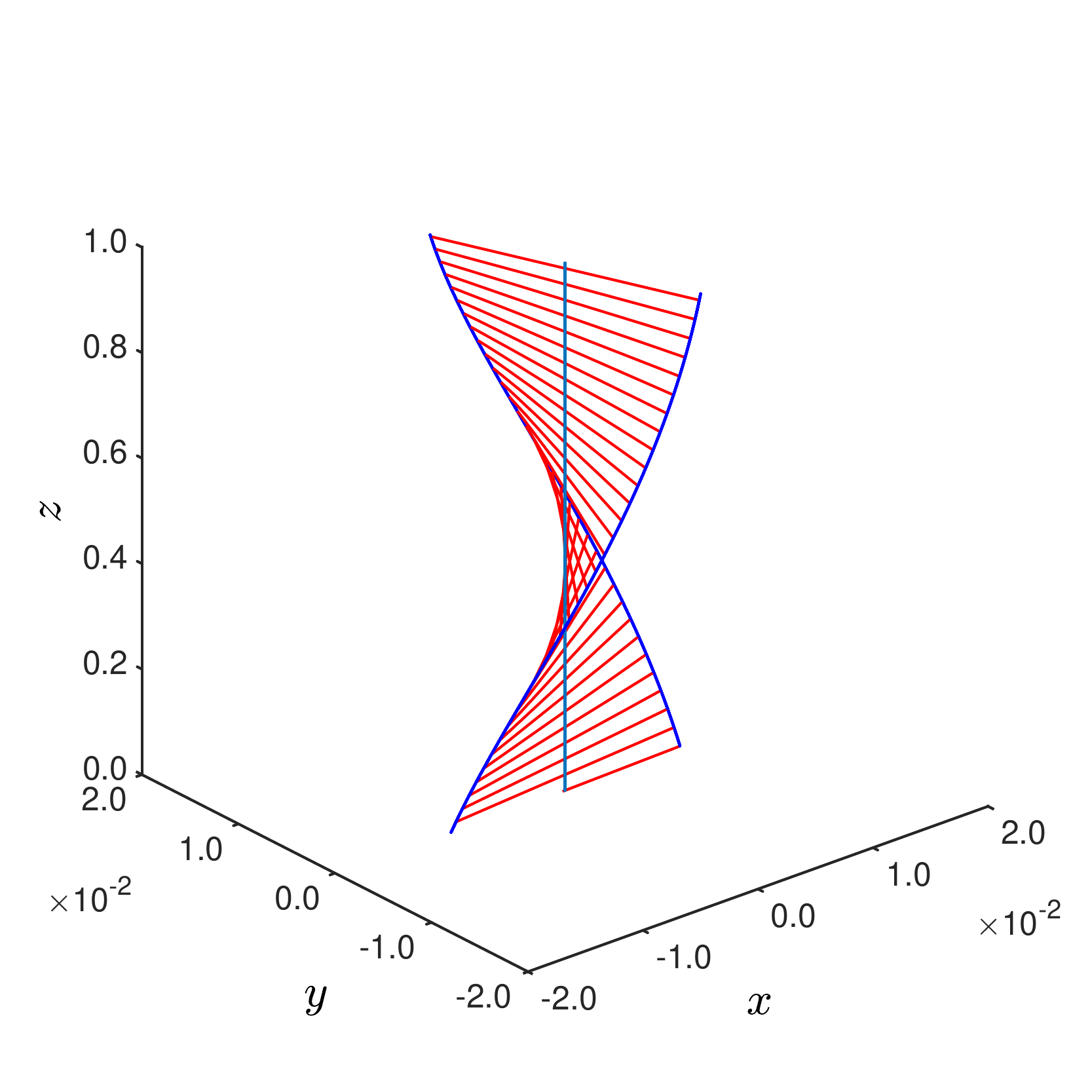} \quad
    \includegraphics[width=0.33\linewidth]{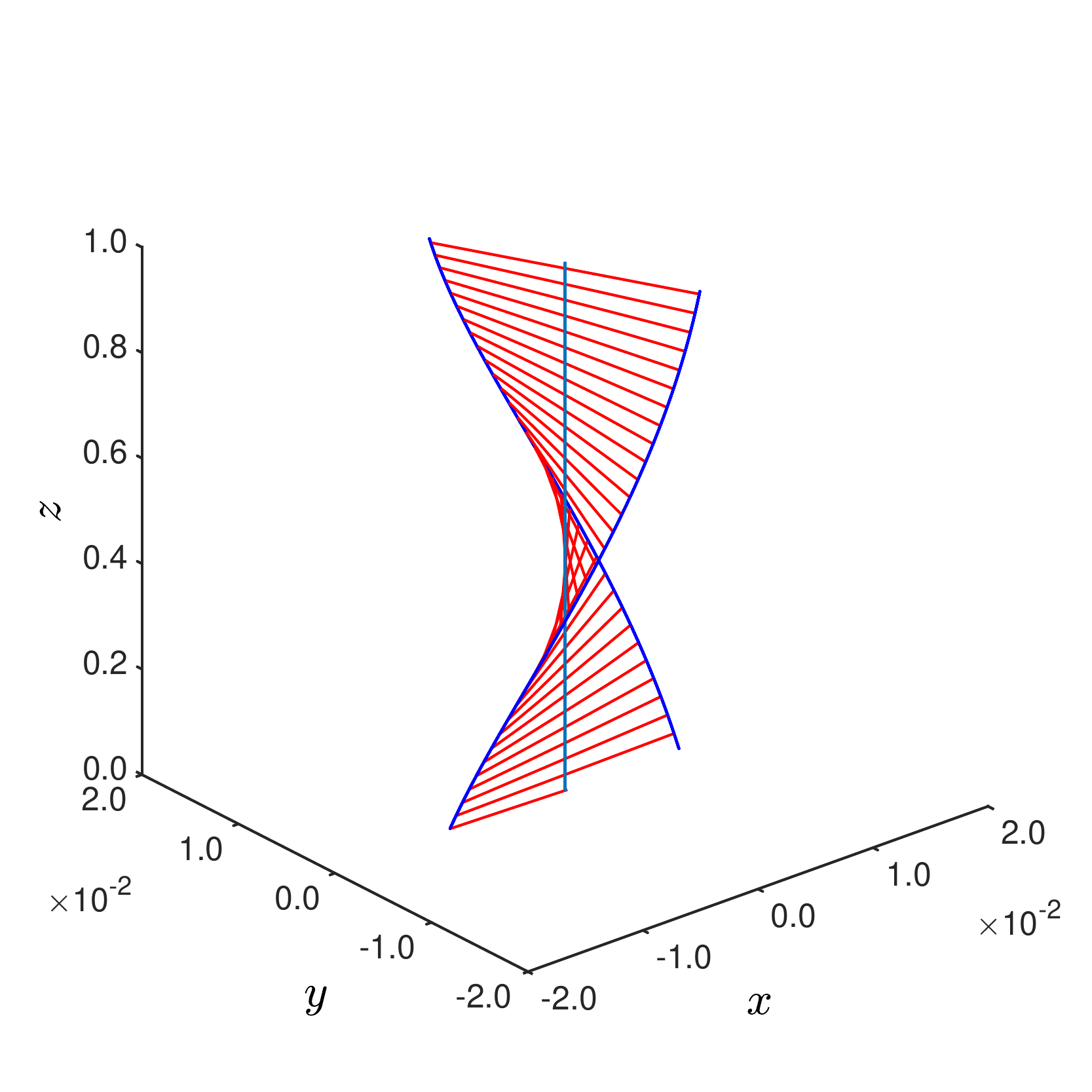}
    \caption{Ribbon with $b_\epsilon = b_H/9$ and a width of $w = 2\times 10^{-2}$, showing a straight twisted solution (helicoid) for $\kappa = 0$ and $\tau = 2$. The aspect ratio has been stretched. The left ribbon was generated with the positive $\epsilon_0$ solution and the right ribbon was generated with the negative $\epsilon_0$ solution. The arc-length step size is $\Delta s = 10^{-4}$, and generators are plotted for every 300 steps.}
    \label{fig:straight}    
\end{figure}

In the limit $\kappa \rightarrow 0$, $\epsilon = \epsilon_0$ as given by~\eqref{eq:ep0}, and we have the following regularized behavior, 
\begin{align}
  \begin{split}
    M_1 &= \frac{\epsilon}{2}(b_H^{1/2}b_\epsilon^{1/2}-3b_\epsilon)
    = \pm\frac{\tau}{2}(b_H^{3/4}b_\epsilon^{1/4}-3b_H^{1/4}b_\epsilon^{3/4})
    \,,
    \\[2mm]
    M_3 &= 2b_H^{1/2}b_\epsilon^{1/2}\,\tau = \pm 2b_H^{1/4}b_\epsilon^{3/4}\epsilon
    \,.
  \end{split}
\end{align}
For $\kappa = 0$, the balance equations are 
\begin{align}
  \begin{split}
    N_1'\pm \left(\frac{b_\epsilon}{b_H}\right)^{1/4} N_2\epsilon &= 0\,,
    \\[2mm]
    N_2' \pm \left(\frac{b_\epsilon}{b_H}\right)^{1/4} N_1\epsilon &= 0\,,
    \\[2mm]
    N_3' &= 0\,,
  \end{split}
  \label{eq:force}
\end{align}
and
\begin{align}
  \begin{split}
    \pm \left(\frac{b_\epsilon}{b_H}\right)^{1/4} M_2\epsilon -N_2 &= 0\,,
    \\[2mm]
    M_2' \pm (b_H^{1/4}b_\epsilon^{3/4}-3b_\epsilon^{5/4}b_H^{-1/4})\frac{\epsilon^2}{2}
    + N_1 &= 0\,,
    \\[2mm]
    \epsilon' &= 0\,.
  \end{split}
  \label{eq:moment4}
\end{align}
The straight twisted rod/helicoidal solution ($M_1 = M_2 = N_1 = N_2 = 0$, $M_3 \sim \tau$, $N_3$ constant) only exists for the parameter ratio $b_\epsilon/b_H = 1/9$. Energy plots suggest that this parameter ratio occurs exactly when we have a transition from a non-convex to a convex energy enveloping the Sadowsky energy. This ratio also confers other desirable behaviors to the model, which will be discussed in the following Section \ref{energetics}. Starting with an orthonormal basis $\{\mathbf{d}_i\}$ at $x = y = z = 0$, based on the proposed kinematics~\eqref{eq:kin}, we plot in Fig.~\ref{fig:straight} this straight rod solution for the strip using $\kappa = 0$, $\tau = 2$, width $w = 2\times10^{-2}$, and an arc-length discretization of $\Delta s = 10^{-4}$. Note that the two $\epsilon_0$~\eqref{eq:ep0} solutions lead to two different embeddings and families of generators. The left image shows the strip generated with the positive $\epsilon_0$ solution, the right with the negative solution. In both figures, generators are plotted for every 300 spatial steps. The difference and associated embedding jump \eqref{eq:jump} are small, proportional to the width of the strip.

If the parameter ratio $b_\epsilon/b_H \neq 1/9$, we no longer have a twisted straight solution with $M_1 = 0$ and $\kappa = 0$. At inflection points, $M_1$ becomes non-zero and suffers a jump $\llbracket M_1\rrbracket = (b_H^{3/4}b_\epsilon^{1/4}-3b_H^{1/4}b_\epsilon^{3/4})\tau$.
 Hence, there is a  \com{discontinuity in the moment} at the inflection point. \com{This concentrated moment is reminiscent of issues observed in ribbon models at inflection points~\cite[Appendix E]{yu2019bifurcations}.}
 \com{The discontinuity} disappears for the parameter ratio $b_\epsilon = b_H/9$.


\section{Energetics}\label{energetics}

We observe that the energy density $e$ from $\mathcal{E}$~\eqref{eq:energy} presents the following limits
\begin{equation}
  e(\kappa,\tau) =
  \begin{cases}
    \dfrac{b_H}{8}\dfrac{(\kappa^2+\tau^2)^2}{\kappa^2} &\text{as}\; \epsilon \rightarrow 0\,,\\[5mm]
    b_\epsilon^{1/2}b_H^{1/2}\tau^2 &\text{as}\; \kappa \rightarrow 0\,,
  \end{cases}       
  \label{eq:elim}
\end{equation}
where the latter was derived using the $\epsilon_0$ solutions from~\eqref{eq:ep0}. The first case holds when $\kappa$ is sufficiently larger than $\tau$, so that $\epsilon = 0$ is the relevant solution for the ribbon, and the energy reduces to the classical Sadowsky functional. 
For example, in the $b_\epsilon / b_H = 1/9$ case discussed in the previous Section~\ref{limits}, the relation~\eqref{eq:zero} indicates that $\epsilon = 0$ for $|\tau| \lesssim 0.43|\kappa|$.
\comb{For any nonzero choice of parameters, the scalings in~\eqref{eq:elim}} connect with the energy density of Freddi and co-workers~\cite{freddi2016corrected},  
\begin{equation}
  e_F(\kappa,\tau) =
  \begin{cases}
    \dfrac{(\kappa^2+\tau^2)^2}{\kappa^2} &\text{if}\; |\kappa| \geq |\tau|\,,\\[5mm]
    4\tau^2 &\text{if}\; |\kappa|\leq |\tau|\,,
  \end{cases}       
  \label{eq:fred}
\end{equation}
where \comb{$\tau = \mathbf{d}_1'\cdot\mathbf{d}_2$,} and all fields depend on $s$. \comb{Both energies  transition from the classical Sadowsky ribbon to a twisted Kirchhoff rod energy quadratic in $\tau$.} \comb{The parameter ratio that would give the same proportionality (prefactors) between the two limits is $b_\epsilon/b_H = 1/4$, for which case the relation~\eqref{eq:zero} indicates that $\epsilon = 0$ for $|\tau| \lesssim 0.62|\kappa|$. 
}

\subsection{Energy landscape}

\begin{figure}[hb]
    \centering
    \begin{subfigure}{0.32\textwidth}    
      \includegraphics[width=1\linewidth]{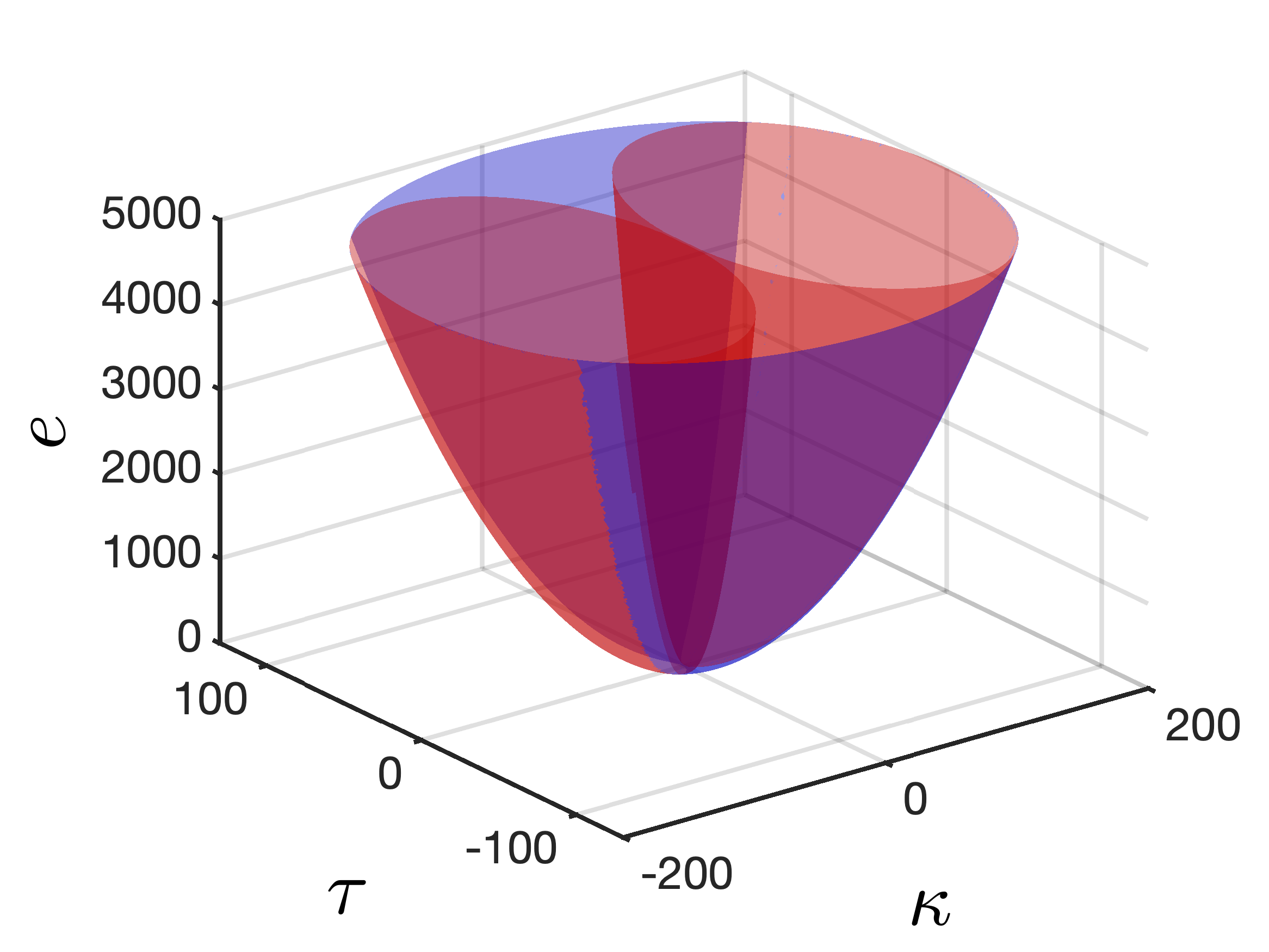} \;
      \caption{Energy density $e$}
    \end{subfigure}
    \begin{subfigure}{0.32\textwidth}    
      \includegraphics[width=1\linewidth]{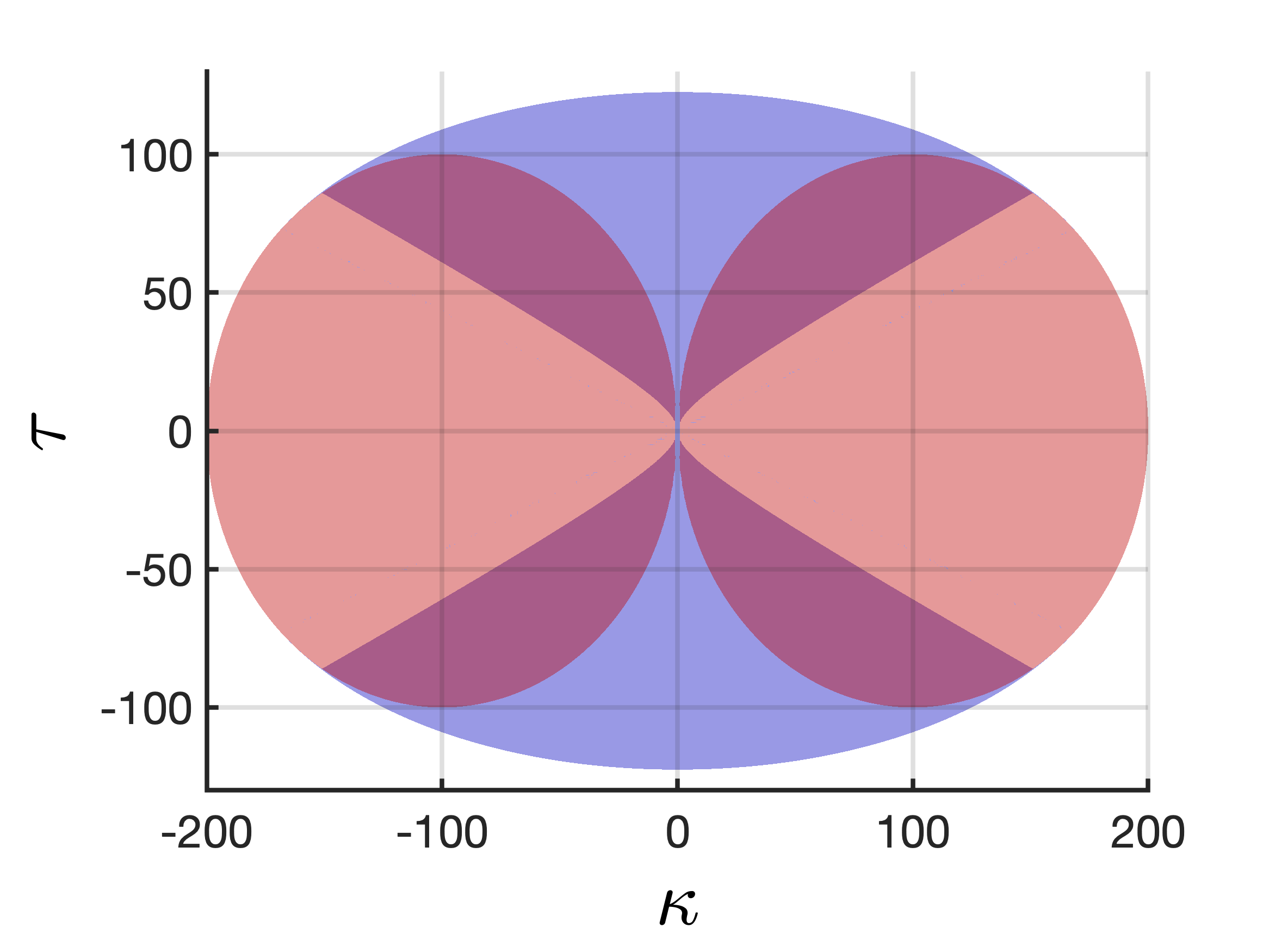} \;
      \caption{$\kappa$-$\tau$ plane view of $e$}
    \end{subfigure}
    \begin{subfigure}{0.3\textwidth}    
    	\begin{center}
      \includegraphics[width=1\linewidth]{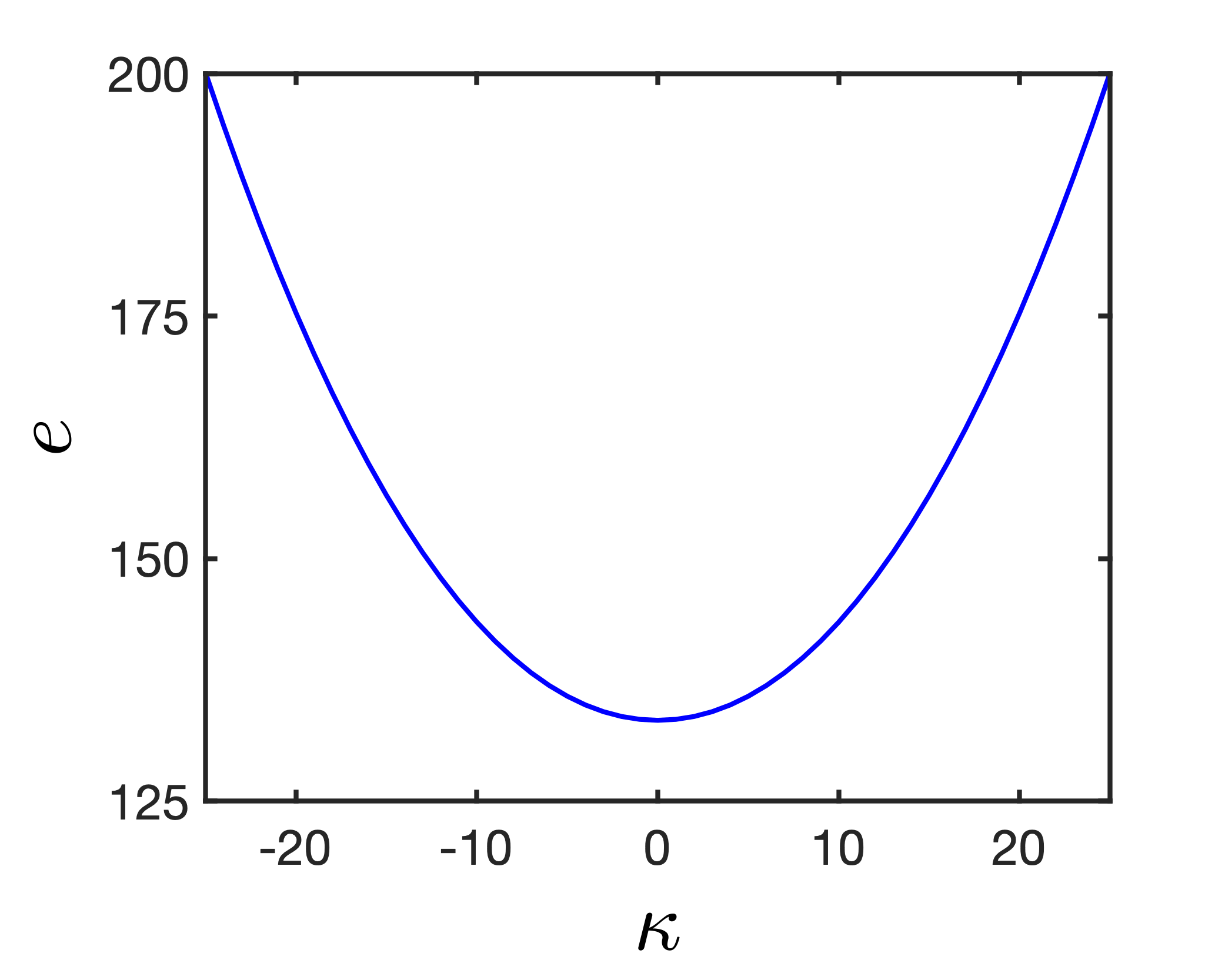}
      \end{center}
      \caption{$e$ profile at $\tau = 20$}
    \end{subfigure}   
    \begin{subfigure}{0.33\textwidth}
      \includegraphics[width=1\linewidth]{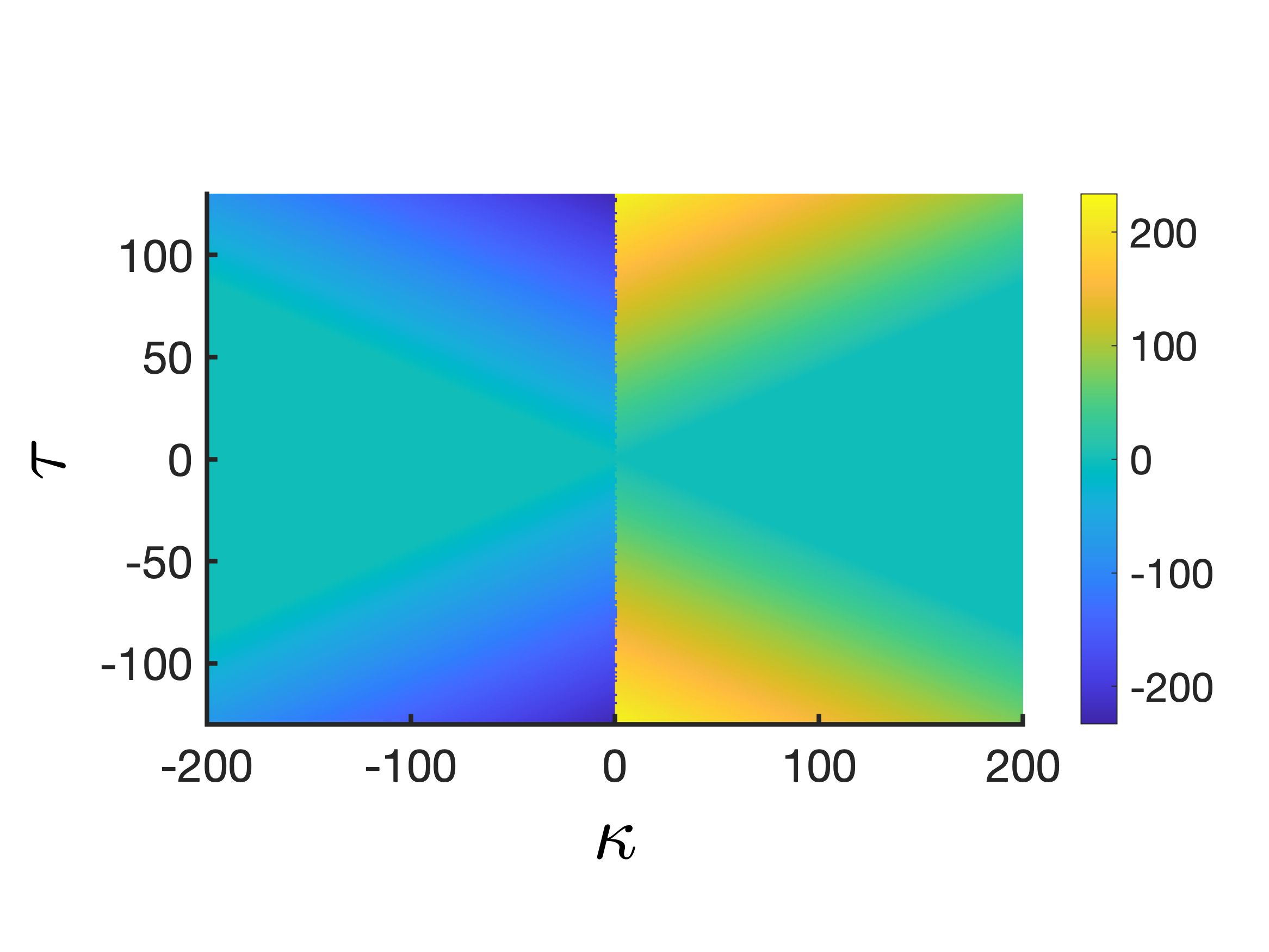} \quad
      \vspace{-6mm}
      \caption{Regularization field $\epsilon$}      
    \end{subfigure}
    \begin{subfigure}{0.27\textwidth}        
      \includegraphics[width=1\linewidth]{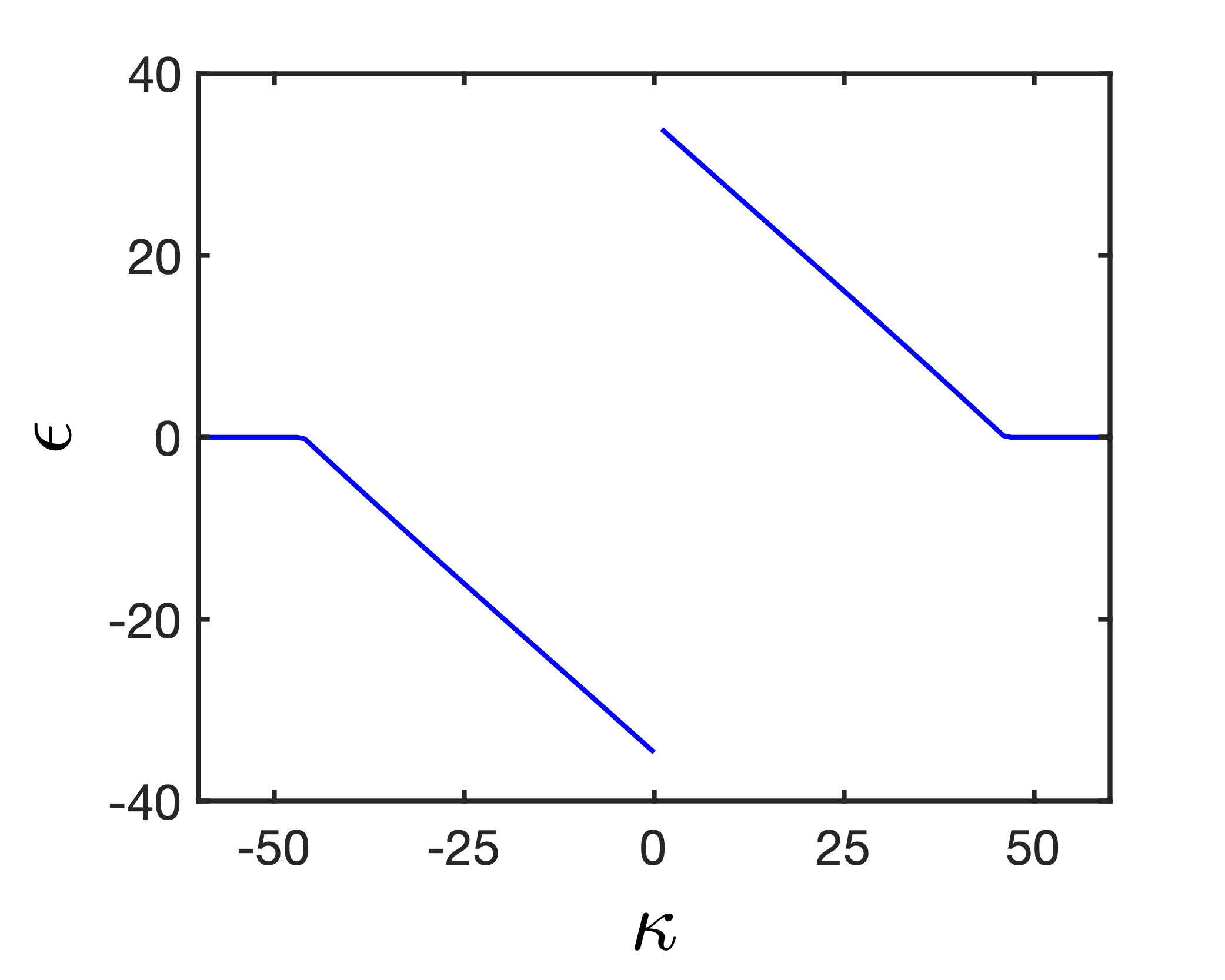}
      \caption{$\epsilon$ profile at $\tau = 20$}
    \end{subfigure}
    \begin{subfigure}{0.27\textwidth}        
      \includegraphics[width=1\linewidth]{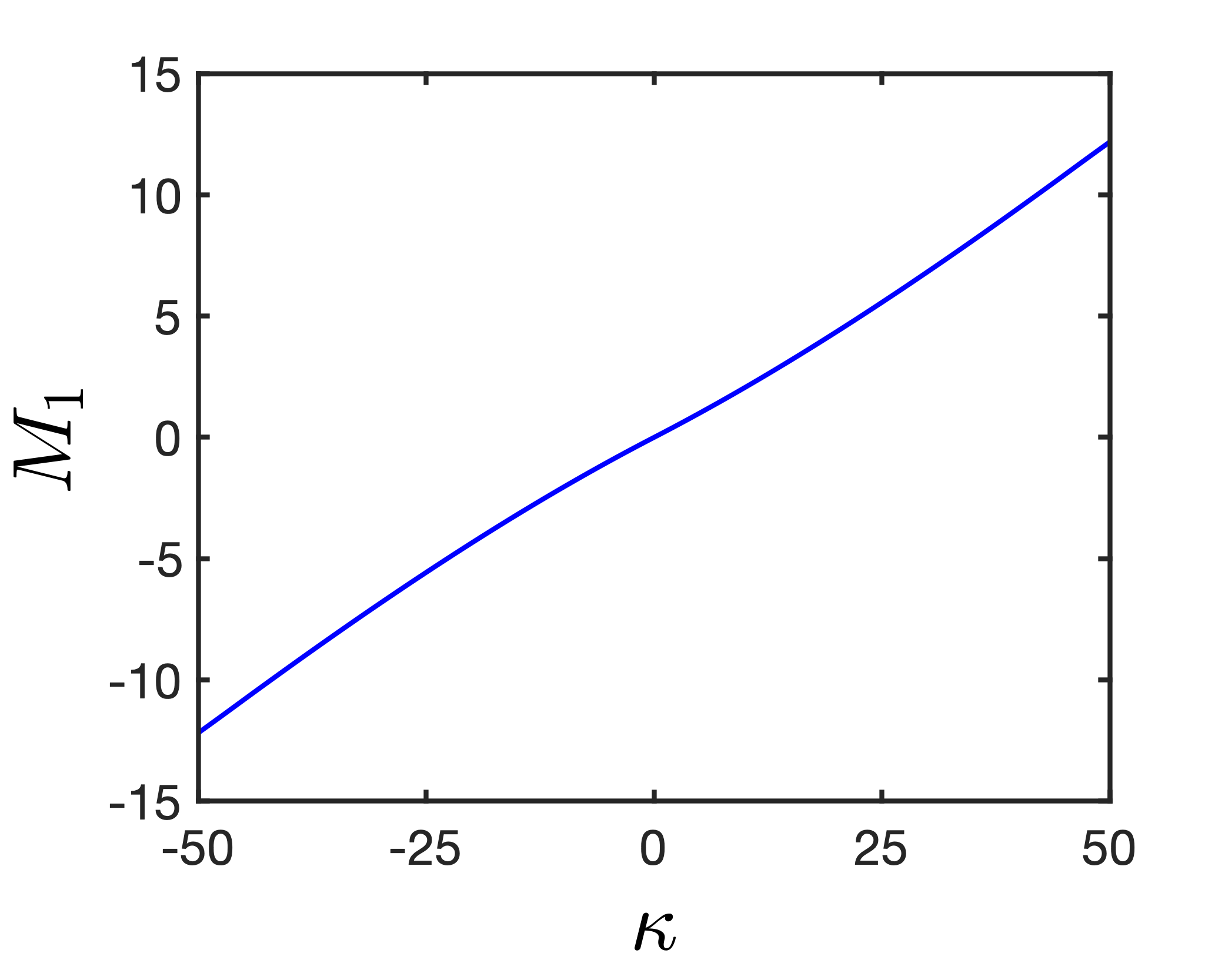}
      \caption{$M_1$ profile at $\tau = 20$}      
    \end{subfigure}
    \caption{Energy density $e$, regularization field $\epsilon$, and moment $M_1$ for $b_\epsilon/b_H = 1/9$.  In (a) and (b) the energy density, as defined in~\eqref{eq:energy}, is shown in blue alongside that of the Sadowsky model in red. Darker colors represent an overlap between the two.}
    \label{fig:energy1}
\end{figure}

\begin{figure}
    \begin{subfigure}{0.32\textwidth}        
      \includegraphics[width=1\linewidth]{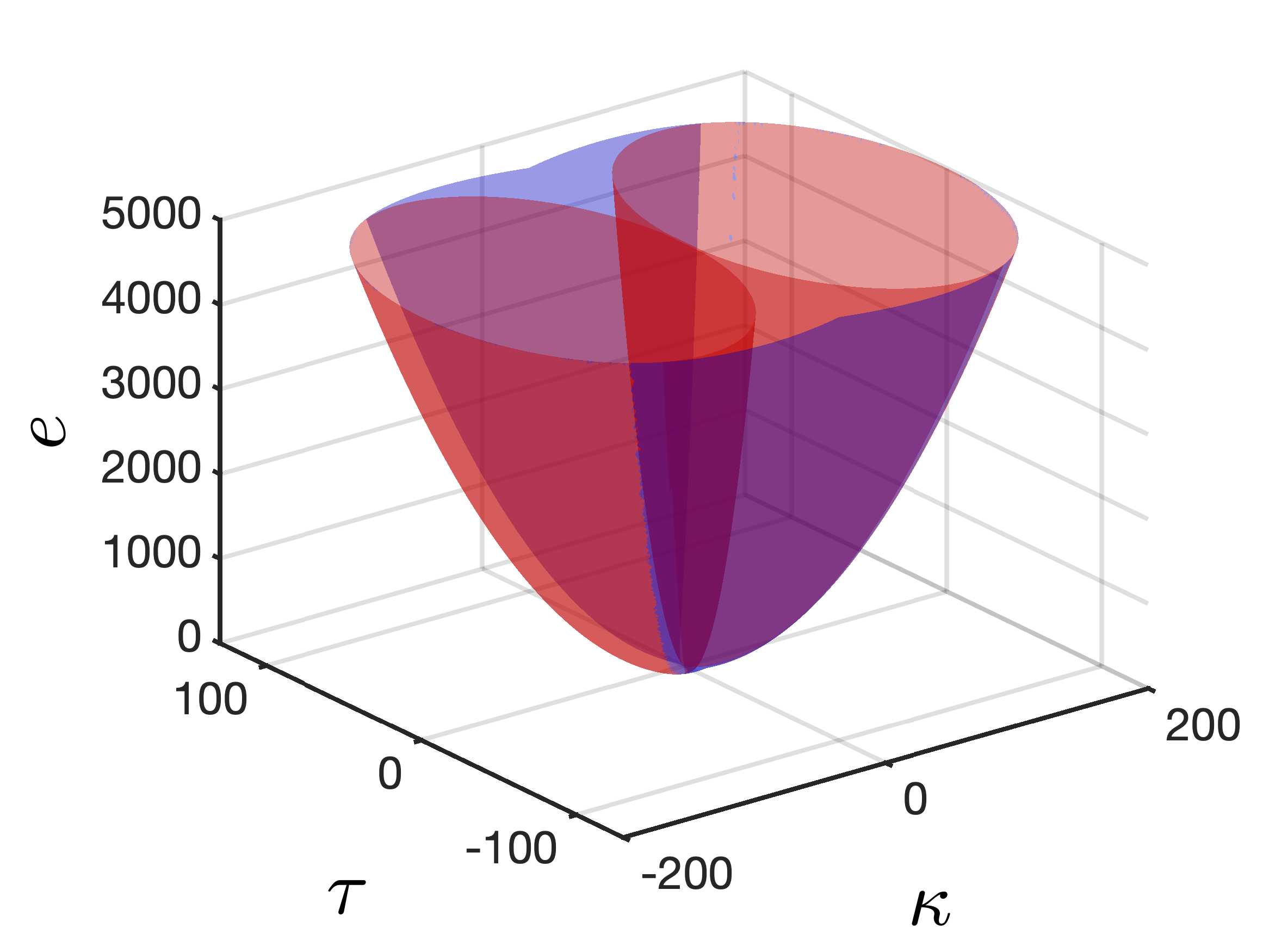} \;
      \caption{Energy density $e$}      
    \end{subfigure}
    \begin{subfigure}{0.32\textwidth}            
      \includegraphics[width=1\linewidth]{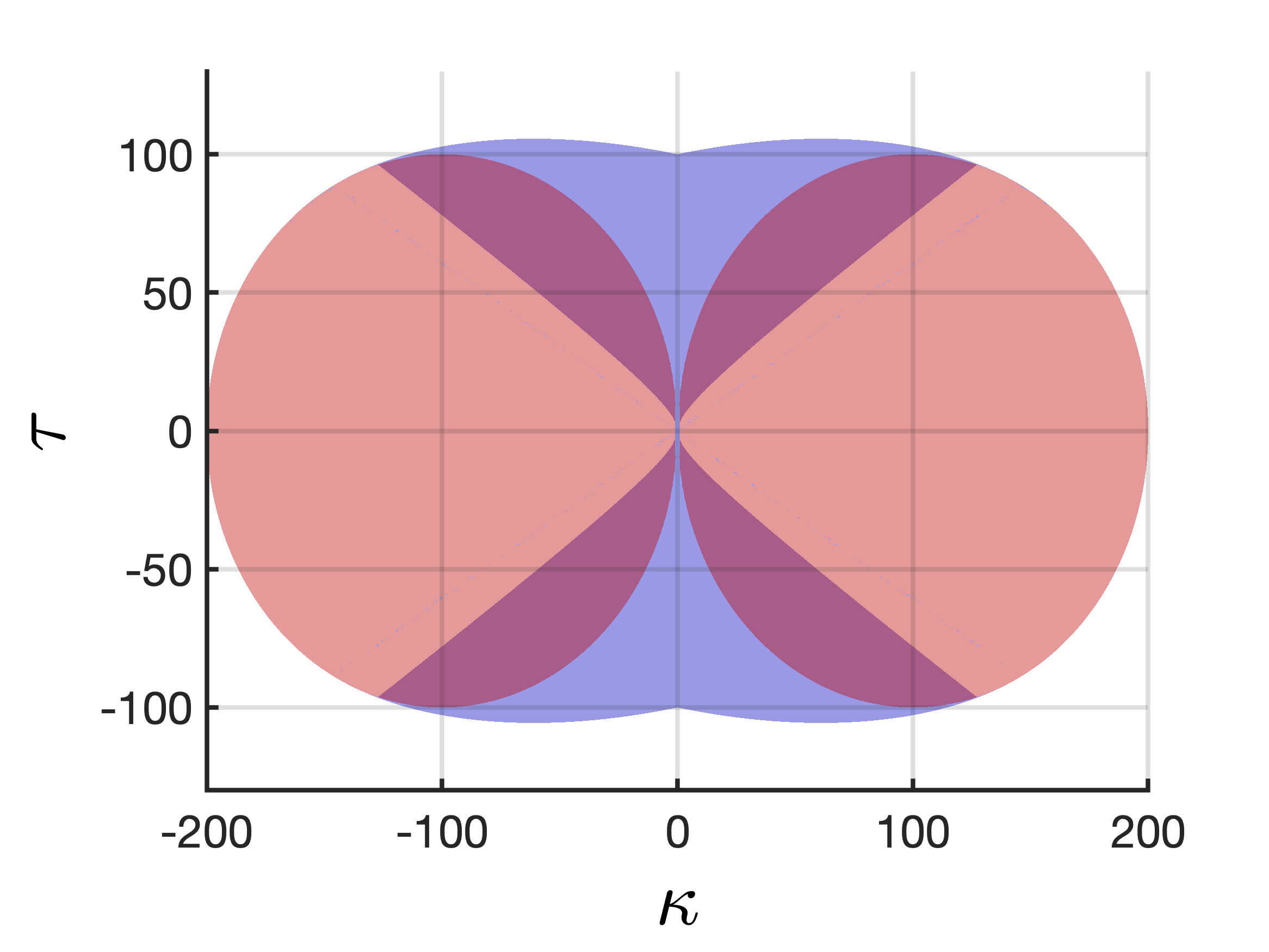} \;
      \caption{$\kappa$-$\tau$ plane view of $e$}      
    \end{subfigure}
    \begin{subfigure}{0.3\textwidth}    
    \begin{center}            
      \includegraphics[width=1\linewidth]{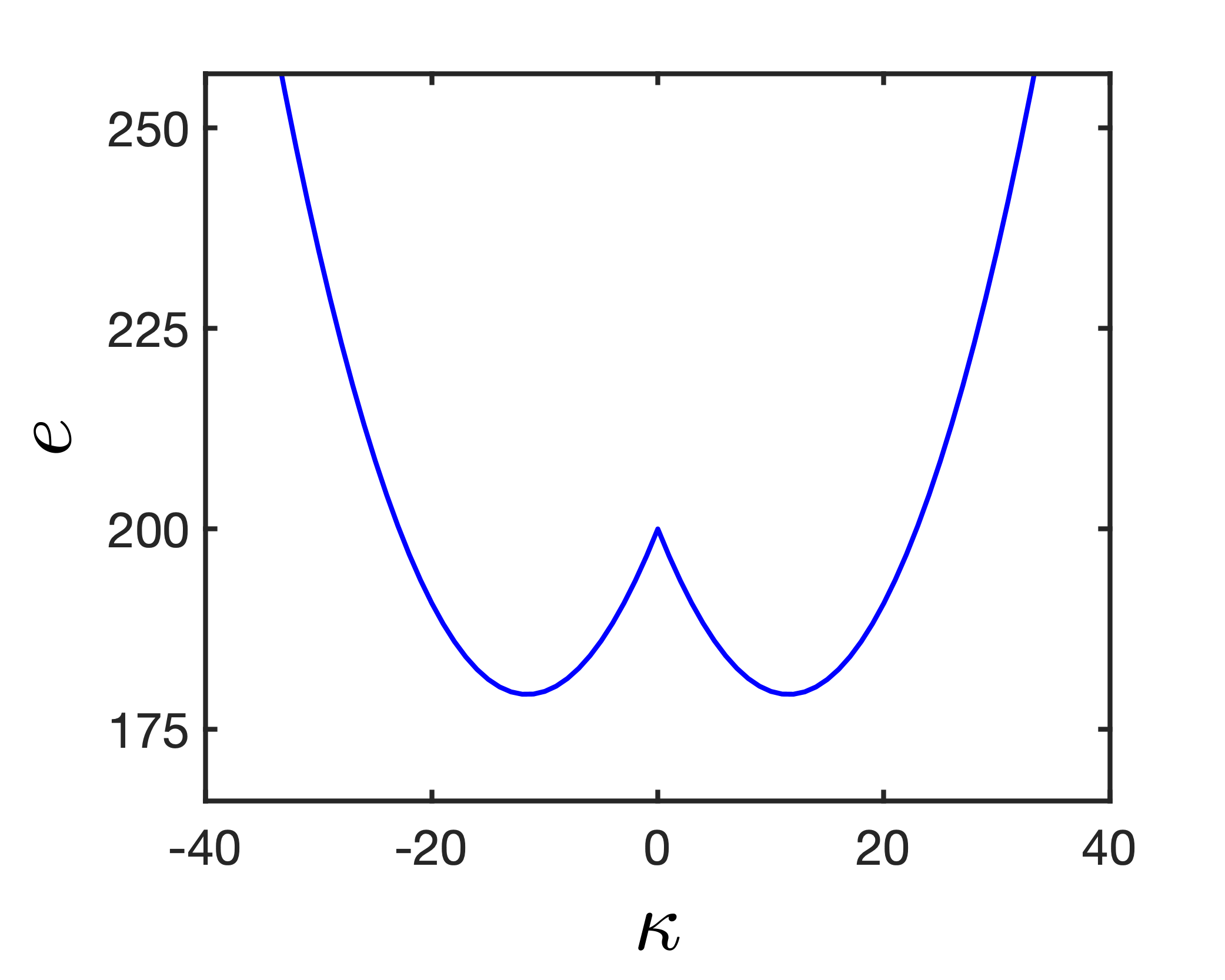}    
      \end{center}
      \caption{$e$ profile at $\tau = 20$}
    \end{subfigure}    
    \begin{subfigure}{0.33\textwidth}        
      \includegraphics[width=1\linewidth]{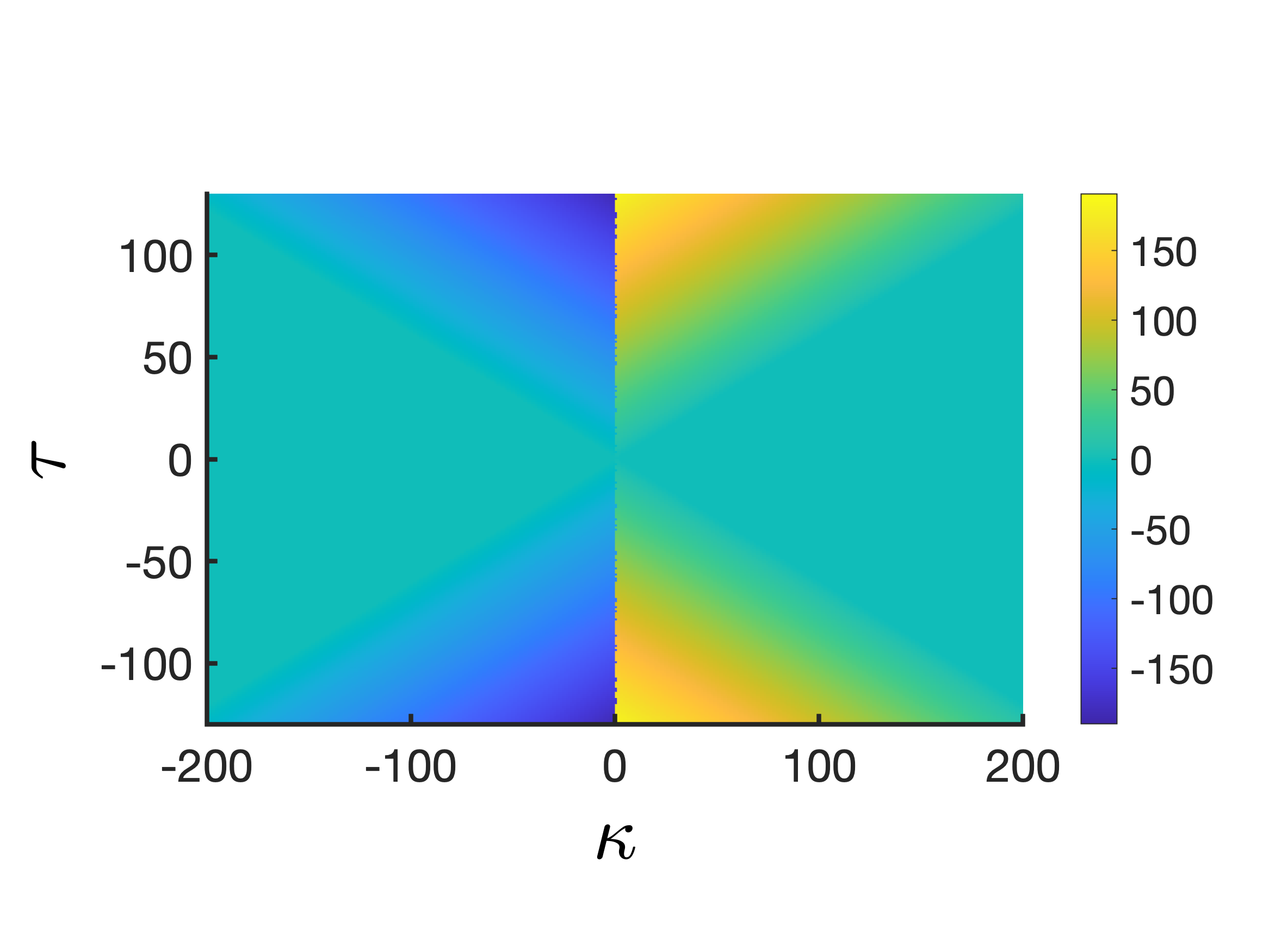} \quad
      \vspace{-6mm}
      \caption{Regularization field $\epsilon$}      
    \end{subfigure}
    \begin{subfigure}{0.27\textwidth}        
      \includegraphics[width=1\linewidth]{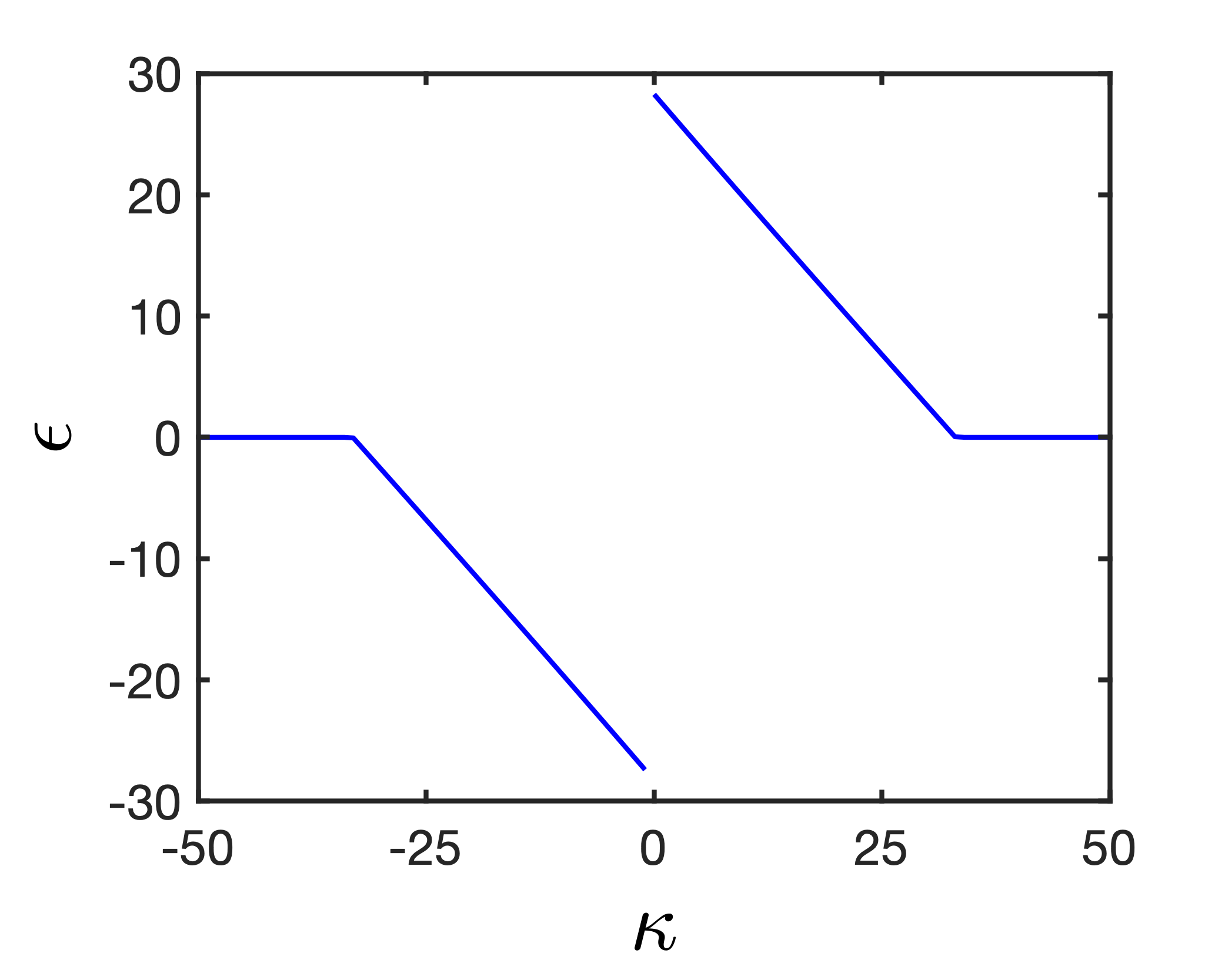}    
      \caption{$\epsilon$ profile at $\tau = 20$}
    \end{subfigure}
    \begin{subfigure}{0.27\textwidth}        
      \includegraphics[width=1\linewidth]{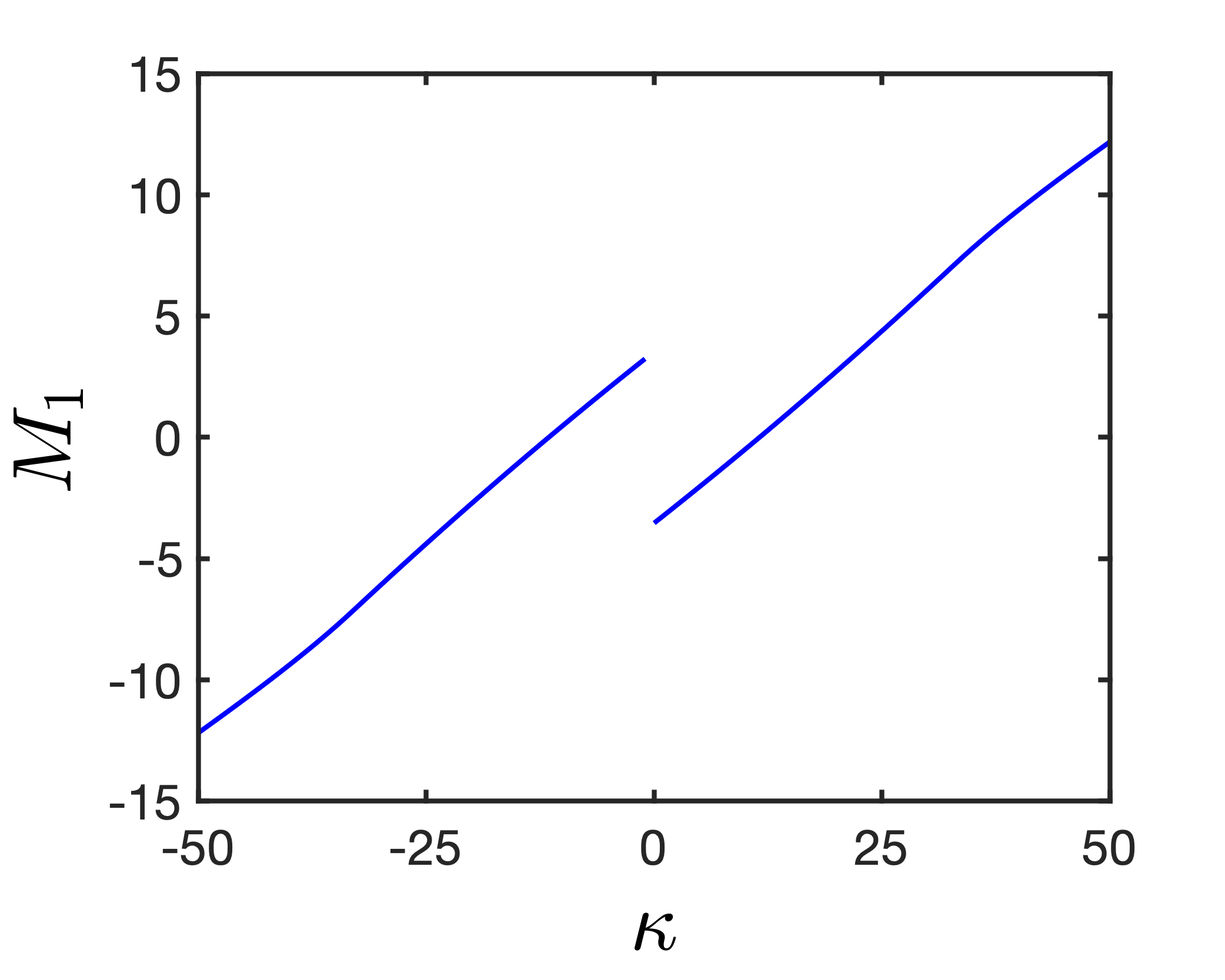}        
      \caption{$M_1$ profile at $\tau = 20$}      
    \end{subfigure}        
    \caption{Energy density $e$, regularization field $\epsilon$, and moment $M_1$ for $b_\epsilon/b_H = 1/4$. In (a) and (b) the energy density, as defined in~\eqref{eq:energy}, is shown in blue alongside that of the Sadowsky model in red. Darker colors represent an overlap between the two.}
    \label{fig:energy2}    
\end{figure}

For a fully representative picture of the energy landscape, we plot respectively in Fig.~\ref{fig:energy1} and Fig.~\ref{fig:energy2} the energy density from~\eqref{eq:energy} for the two cases of interest we have identified so far: \com{the convex $b_\epsilon/b_H = 1/9$ case for which there is no jump in the moment at inflection points but the prefactors do not match~\eqref{eq:fred}, and the $b_\epsilon/b_H = 1/4$ case for which the prefactors match~\eqref{eq:fred} but there is a jump in the moment.}  We also plot the Sadowsky energy, which takes the form of two separate wells.  

\begin{figure}[ht]
    \centering
    \includegraphics[width=0.38\linewidth]{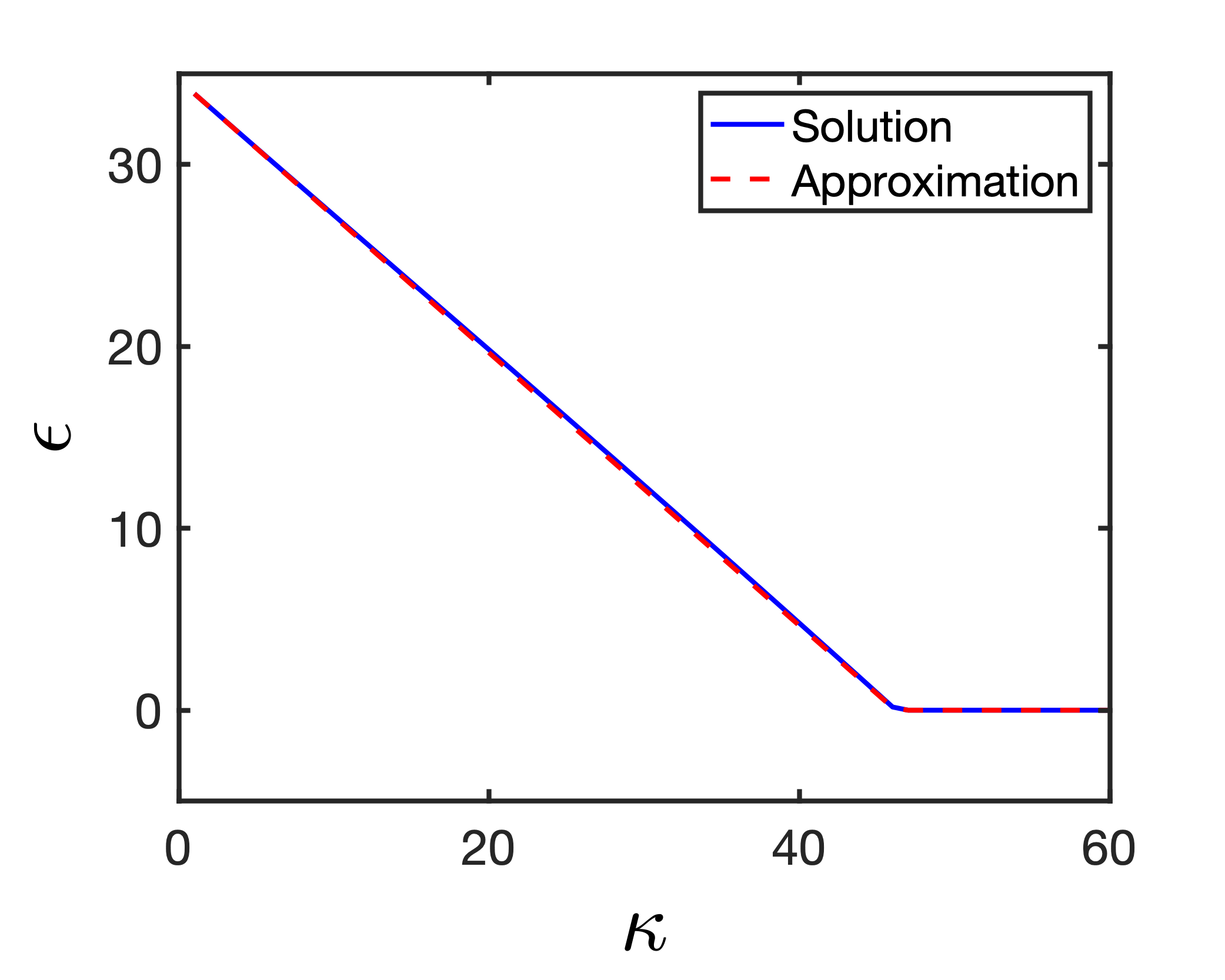} \quad
    \caption{
    Comparison of the linear approximation \eqref{eq:epasym} (dashed red) and exact solution \eqref{eq:ep} (solid blue) of the regularization field $\epsilon$ over part of the constant-$\tau$ slice shown in Figure \ref{fig:energy1}(e).
    }
    \label{fig:approx}    
\end{figure}

As shown in Fig.~\ref{fig:energy1}, the energy for $b_\epsilon/b_H = 1/9$ is convex, allowing for a single $(\kappa,\tau)$ solution in the large $\tau/\kappa$ limit. As shown in Fig.~\ref{fig:energy2}, the energy for $b_\epsilon/b_H = 1/4$ is non-convex, presenting two overlapping wells in $\kappa$ for a fixed value of $\tau$. 
 When $|\tau| > |\kappa|$, non-convex models such as this, or the non-convexified model from~\cite{audoly2021one}, are unstable to the formation of ``microstructure''-- decomposition into a mixture of two $\kappa$ values for the same $\tau$ value. 
The $\epsilon$ fields obtained from solving~\eqref{eq:ep} are very well approximated by the form~\eqref{eq:epasym}, as shown in Fig.~\ref{fig:approx} for a slice through the $b_\epsilon/b_H = 1/9$ case.

\subsection{A note about the moment $M_1$}

The moment component $M_1$ suffers a jump at inflection points for all but a special choice of parameter ratio. 
As can be gleaned from Figures \ref{fig:energy1} and \ref{fig:energy2}, its behavior is otherwise approximately linear in $\kappa$ for any $(b_\epsilon,b_H)$ combination.  
With regard to jumps, we identify three behaviors:
\begin{itemize}
\item $b_\epsilon/b_H = 1/9$: $M_1 = 0$ at $\kappa = 0$, and a straight twisted rod solution is obtained for $\tau \neq 0$.
\item $b_\epsilon/b_H > 1/9$: $M_1$ has a ``negative'' jump at $\kappa = 0$, with $M_1 = 0$ at the two energy minima in $\kappa$, as in Figure \ref{fig:energy2}. The twisted strip can present alternating $\pm\kappa$ regions, like the triangulated solution discussed in~\cite{audoly2023analysis}.
\item $b_\epsilon/b_H < 1/9$: $M_1$ never crosses zero, and has a ``positive'' jump at $\kappa = 0$. 
\end{itemize}

\subsection{Previous works}\label{comparison}

We now briefly describe some of the previously proposed ribbon models that explored regularization by relaxing the inextensibility constraint~\cite{ghafouri2005helicoid,sano2019twist,audoly2021one,kumar2025centerline}. One application of particular interest to these authors is the helicoid to spiral-ribbon transition. Helicoids are twisted ribbons with zero curvature $\kappa$, with a Gaussian curvature that scales as $K \sim -\tau^2$, whereas spiral ribbons are curved ribbons with zero Gaussian curvature (see, for example, \cite{ghafouri2005helicoid}). The energetics of each of these regimes is captured by the piecewise Freddi energy \cite{freddi2016corrected}.

Ghafouri and Bruinsma~\cite{ghafouri2005helicoid} 
proposed, and Sano and Wada~\cite{sano2019twist} later adopted, a regularization of the Sadowsky energy employing a small positive constant $C$ in the form $\tau^4/(C+\kappa^2)$.
 The problem with this approach is that the ribbon energy has a scaling of $\sim\tau^4$ when twist dominates over curvature, which does not match the {Freddi et al. energy scaling}~\cite{freddi2016corrected} of $\tau^2$ in this regime.
 Moreover, the Gaussian curvature $K$ becomes a function of $\kappa$ and $\tau$, which is proportional to the small regularization term, and nonzero for $\tau\neq 0$. While these theories correctly give $K \sim -\tau^2$ for helicoids, their spiral ribbons have a small but nonzero $K$ when $\tau \neq 0$; this poses a problem to the Gauss-Codazzi equations~\cite{sano2019twist}, as embedding of such ribbons requires $K=0$.
 Authors interested in exploring twisted to spiral ribbon transitions have proposed alternative approaches that circumvent some of these problems \cite{grossman2016elasticity,sawa2011shape,teresi2013modeling,tomassetti2017capturing,agostiniani2017shape}. 

The corrected Sadowsky functional~\eqref{eq:fred} derived by Freddi and co-workers~\cite{freddi2016corrected,freddi2016variational} addressed these issues. For example, when $\tau$ is larger than $\kappa$, their piecewise energy captures the proper scaling for a helicoidal configuration, and permits transitions between helicoidal to spiral-ribbon configurations~\cite{barsotti2022stability, barsotti2019straight}. 
In~\cite{neukirch2021convenient}, Neukirch and Audoly adopted this energy and derived constitutive expressions for $\kappa$ and $\tau$, seeking a convenient formulation for Sadowsky. To do this, they inverted the constitutive relations for the moments $M_2$ and $M_3$. However, this introduced a discontinuity in the model when $M_2 = 0$ and $M_3 \neq 0$. The authors addressed this by regularizing the constitutive relation for $\kappa$ through the introduction of an artificial constant, which effectively changes the ribbon model. In~\cite{audoly2021one}, Audoly and Neukirch derived a 1D ribbon model from a nonlinear plate model. Their treatment leads to a function of the curvature that represents the effects of stretching away from the centerline, and regularizes the energy in the limit of vanishing curvature.
While this model recovers Sadowsky for $\kappa \gg 1$, and the Kirchhoff rod for small $\kappa$ and $\tau$, it does not match the Freddi energy scaling of $\tau^2$~\cite{freddi2016corrected} for general $|\tau/\kappa| > 1$. Their energy is non-convex, and solutions can feature discontinuous distribution of $\kappa$ and $\tau$ (microstructure). They further explored how patching the energy could turn it into a convexified Sadowsky functional connected with the Freddi energy. 

The model of Kumar and Rangarajan \cite{kumar2025centerline} is also derived from a plate model, but retains an auxiliary field representing transverse curvatures, as in our direct model.  As mentioned briefly in our introduction, we are unable to ascertain the energetic scalings of this model without further information about the behavior of solutions of their auxiliary field.

\section{Example: Spiral-ribbon states}

As an example, we look at spiral-ribbon states characterized by constant $\kappa$ and $\tau$. This analysis is motivated by the work of Barsotti and co-workers ~\cite{barsotti2022stability}, who studied a 1D ribbon model governed by the corrected Sadowsky energy~\eqref{eq:fred} with a material frame that followed the Frenet-Serret kinematics~\eqref{eq:kin}. Based on that energy, when twist prevails ($|\tau| \geq |\kappa|$), a Sadowsky spiral-ribbon transitions into a ``frustrated'' spiral-ribbon state, and when $\kappa$ vanishes, the latter transitions into a frustrated helicoid.  We perform a similar analysis to compare with the spiral-ribbon configurations and the internal force of a ribbon governed by our model.

Define $\{\mathbf{e}_i\}$ as a particular choice of the reference orthonormal $\{\mathbf{D}_i\}$ basis, associated with an undeformed straight ribbon. In this case, $\mathbf{e}_3$ is tangent to the straight curve, $\mathbf{e}_2$ is the normal, and $\mathbf{e}_1 = \mathbf{e}_2\times\mathbf{e}_3$. Assume the ribbon deformation can be represented by a curve $\mathbf{x}(s)$ with a set of directors satisfying~\eqref{eq:kin}, while enforcing the $\kappa_2=0$ constraint. If $\kappa$ and $\tau$ are constant, and $\kappa \geq 0$ and $\kappa^2+\tau^2 > 0$, it can be shown~\cite{docarmo1976differential,barsotti2022stability} that, up to a rigid displacement, the equation for the curve is given by 
\begin{equation}
    \mathbf{x}(s) = \lambda s\,\mathbf{e}_3-\rho\sin(\gamma s) \mathbf{e}_1+\rho\cos(\gamma s)\mathbf{e}_2 \,,
    \label{eq:spiral}
\end{equation}
where
\begin{equation}
    \lambda = \sqrt{\frac{\tau^2}{\kappa^2+\tau^2}}\;,\;
    \rho = \frac{\kappa}{\kappa^2+\tau^2}\;,\;
    \gamma=\pm\sqrt{\kappa^2+\tau^2} \,.
\end{equation}
Observe that $\lambda^2+\gamma^2\rho^2 = 1$, and hence $\kappa = \rho\gamma^2$ and $\tau = \lambda\gamma$. The normal to the curve is given by
\begin{equation}
\mathbf{d}_2=-\sin(\gamma s)\mathbf{e}_1+\cos(\gamma s)\mathbf{e}_2 \,.
\label{eq:spiraln}
\end{equation}
We will focus our attention on the $\kappa \geq 0$, $\tau \geq 0$, and $\epsilon \geq 0$ case.

In the undeformed state, the ribbon presents $\gamma = 0$. If $\gamma > 0 $ but $\rho = 0$, then the ribbon is in the helicoidal configuration shown in Fig.~\ref{fig:straight}, with $\lambda = 1$. Finally, if $\gamma > 0$ and $\lambda < 1$, then the ribbon is in a spiral-ribbon state. For $\gamma > 0$, the internal force and moment of the~\eqref{eq:spiral} configuration can be written as follows~\cite{barsotti2022stability}:
\begin{equation}
    \mathbf{N} = N_\epsilon \mathbf{e}_3
    \quad,\quad
    \mathbf{M} = M_\epsilon\mathbf{e}_3-\rho N_\epsilon\mathbf{d}_2\times\mathbf{e}_3 \,.
\end{equation}
Based on~\eqref{eq:spiraln}, we identify the following expression for the component $N_\epsilon$:
\begin{equation}
    N_\epsilon = \lambda N_3+\rho\gamma N_1 \,.
\end{equation}
Since $\kappa$ and $\tau$ are constant, the components $N_1$ and $N_3$ can be determined from the equilibrium equations~\eqref{eq:force2} and~\eqref{eq:moment2} as
\begin{equation}
    N_1 = M_3\kappa-M_1\tau \;,\quad N_3 = \frac{\tau}{\kappa} N_1.
    \label{eq:spiralf}
\end{equation}
However, in the helicoid state ($\lambda = 1$), $N_3$ is an undetermined constant. Based on~\eqref{eq:spiralf}, for $\lambda < 1$, the component $N_\epsilon$ becomes
\begin{equation}
    N_\epsilon = \frac{N_1}{\rho\gamma} \,.\label{eq:neps}
\end{equation}

\begin{figure}[ht]
    \centering
    \includegraphics[width=0.6\linewidth]{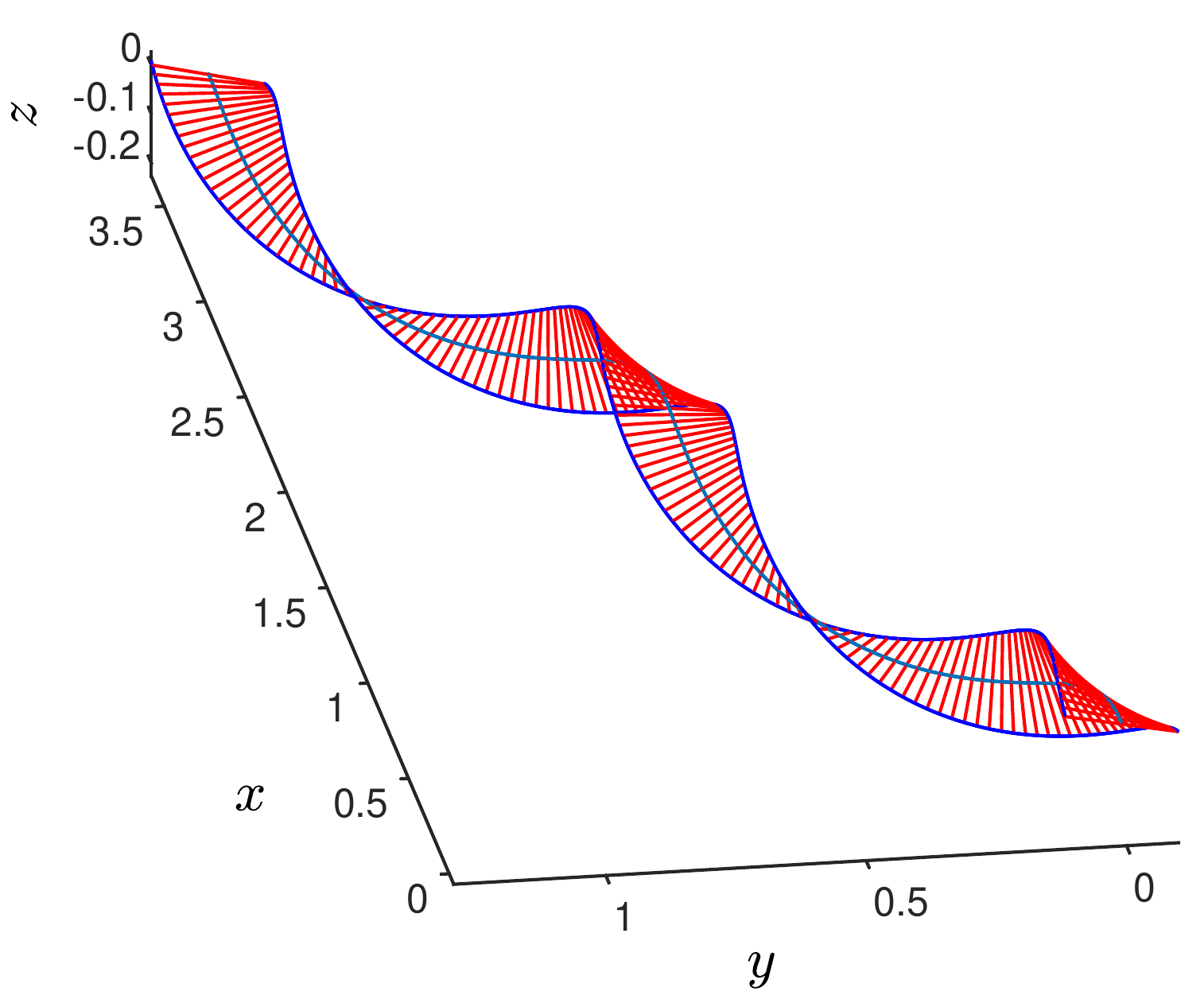}
    \caption{Ribbon with $b_\epsilon = b_H/9$ and a width of $w = 2\times 10^{-1}$, showing a spiral-ribbon of constant $\kappa = 1$ and $\tau = 3$. The ribbon was plotted using~\eqref{eq:f} and~\eqref{eq:spiral} with the $\epsilon$ solution from~\eqref{eq:ep}. The arc-length step size is $\Delta s = 10^{-4}$, and generators are plotted for every 300 steps.}
    \label{fig:spiral1}    
\end{figure}

Set the ratio $b_\epsilon/b_H = 1/9$. When $\epsilon = 0$, the internal force $N_\epsilon$ behaves exactly like the one of a Sadowsky spiral-ribbon. Based on~\eqref{eq:zero}, above $\lambda \approx 0.395$ (i.e. above $\tau \approx 0.43\, \kappa$) we find $\epsilon > 0$. In this case, the ribbon develops a constant Gaussian curvature $K=-\epsilon^2f^2$. This deviates from the classic $K=0$ spiral-ribbon, so we refer to this configuration as a ``curled spiral-ribbon". Therefore, when $\lambda$ is above this value, $N_\epsilon$ is neither a Sadowsky spiral-ribbon nor the frustrated spiral-ribbon of~\cite{barsotti2022stability}. To illustrate, in Fig.~\ref{fig:spiral1} we plot the configuration described by~\eqref{eq:spiral} using $\tau = 3$ and $\kappa = 1$, so that $\lambda = 0.9487$. The spiral-ribbon is constructed using~\eqref{embedding} with $\epsilon$ obtained from~\eqref{eq:ep}. Despite the vanishing width limit of Sadowsky, we use an exaggerated width value of $w = 2\times10^{-1}$ to clearly show the effect of $\epsilon$ on embedding. The curled spiral-ribbon state bridges the $K=0$ classic spiral-ribbon state to the $K\sim-\tau^2$ helicoid state.

In contrast to \eqref{eq:neps}, the corresponding term $N_c$ identified by~\cite{barsotti2022stability} has the piecewise nature of the corrected energy~\eqref{eq:fred},
\begin{equation}
  N_c(\kappa,\tau) =
  \begin{cases}
    \dfrac{S\gamma^2\lambda}{(1-\lambda^2)^2} &\text{if}\; 0 \leq \lambda \leq \frac{1}{\sqrt{2}}
    \\[3mm]
    4S\gamma^2\lambda &\text{if}\; \frac{1}{\sqrt{2}} \leq \lambda < 1 \,.
  \end{cases}       
  \label{eq:nfred}
\end{equation}
They observed that the transition from a Sadowsky spiral-ribbon to a frustrated spiral-ribbon ($\frac{1}{\sqrt{2}} \leq \lambda < 1$) was smooth except for the derivative of the term $N_c$ with respect to $\lambda$.

\begin{figure}[ht]
    \centering
    \includegraphics[width=0.6\linewidth]{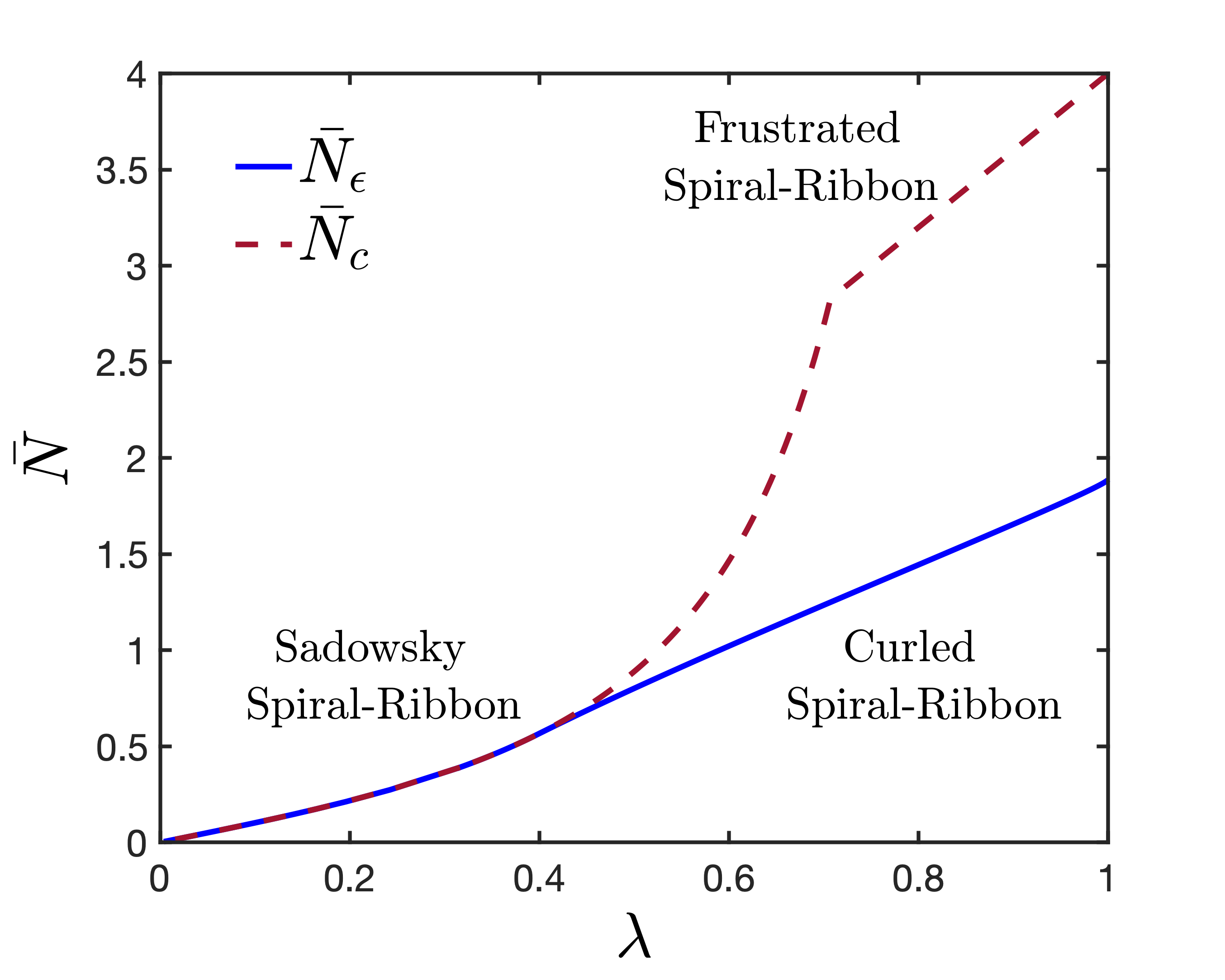} \quad
    \caption{
    Spiral-ribbon equilibrium states as a function of $\lambda$. These are special solutions of constant curvature $\kappa$ and twist $\tau$. $\bar{N}_\epsilon = 4N_\epsilon/(b_H\gamma^2)$ is a scaled force derived from our regularized model~\eqref{eq:energy}, whereas $\bar{N}_c=N_c/(S\gamma^2)$ is a scaled force derived from~\eqref{eq:fred}, which was investigated by Barsotti et al.~\cite{barsotti2022stability}.
    }
    \label{fig:spiral2}    
\end{figure}

In Fig.~\ref{fig:spiral2} we compare the behavior of $\bar{N}_\epsilon = 4N_\epsilon/(b_H\gamma^2)$ against $\bar{N}_c=N_c/(S\gamma^2)$ as a function of $\lambda$. For $\bar{N}_\epsilon$, we observe a smooth transition between the Sadowsky and curled spiral-ribbon states at $\lambda \approx 0.395$. In addition, the $\bar{N}_\epsilon$ curve in the curled spiral-ribbon state suggests a linear behavior as a function of $\lambda$. Note that the $\epsilon$ solution obtained from~\eqref{eq:ep} corresponds to the value associated with the convex energy from Fig.~\ref{fig:energy1}. Similarly to the frustrated spiral-ribbon, the transition from the curled spiral-ribbon to the helicoid state at $\lambda=1$ may present a discontinuity in the force since $N_3$ is undetermined at that point.

\section{Perturbation expansion around an inflection point.}\label{perturbation}

We know there will be a discontinuity of the regularization field $\epsilon$ at a point of inflection, and seek to investigate how this affects physical variables in the vicinity, in the limit of vanishing width. We thus conduct a perturbation expansion in the limit of small $\kappa$, which considers the solutions above (denoted $(\cdot)^+$) and below (denoted $(\cdot)^-$) an inflection point at $s=0$, in the absence of external forces and torques.  We seek to enforce continuity of physical fields at the inflection point, and elsewhere assume that all functions are suitably differentiable. 

At the inflection point $s=0$, $\kappa(0) =0$ and $\tau(0) =\tau_0 \neq 0$. Hence we look for solutions of the form
\begin{eqnarray}
	\begin{split}
		\kappa^\pm &=& 0 +  \varepsilon \kappa^{(1)\pm}(s) + \dots, \\
		\tau^\pm &=& \tau^{(0)\pm}(s) +   \varepsilon \tau^{(1)\pm}(s) + \dots, \\
		\epsilon^\pm &=& \epsilon^{(0)\pm}(s) + \varepsilon  \epsilon^{(1)\pm}(s) + \dots, \\
		b_\epsilon^\pm &=& b_{\epsilon}^{(0)\pm} + \varepsilon b_{\epsilon}^{(1)\pm} + \dots, \\ 
		N_1^\pm &=& N_1^{(0)\pm}(s) + \varepsilon  N_1^{(1)\pm}(s) + \dots, \\
		N_2^\pm &=& N_2^{(0)\pm}(s) + \varepsilon N_2^{(1)\pm}(s) + \dots, \\
		N_3^\pm &=& N_3^{(0)\pm}(s) + \varepsilon N_3^{(1)\pm}(s) + \dots, \\
		M_2^\pm &=& M_2^{(0)\pm}(s) + \varepsilon  M_2^{(1)\pm}(s) + \dots, 
	\end{split}
\end{eqnarray}
where $\varepsilon$ is a small parameter related to the scale of the curvature,  
and $b_\epsilon$ is a constant, while all other quantities are fields depending on $s$. 
Recall that $b_\epsilon/b_H = 1/9$ was suggested by our earlier investigations.   
In this expanded form the mean curvature (Eq.~\eqref{eq:gc}) and the first and third moments (from Eqs.~\eqref{eq:moment1}) become
	\begin{eqnarray}
		\begin{split}
		H &=& - \frac{(\tau^{(0)\pm})^2}{\epsilon^{(0)\pm}} + \varepsilon \frac{(\tau^{(0)\pm})^2 (2 \epsilon^{(1)\pm} +3 \kappa^{(1)\pm}) - \epsilon^{(0)\pm} (4 \tau^{(0)\pm} \tau^{(1)\pm}+\epsilon^{(0)\pm}) }{2 (\epsilon^{(0)\pm})^2} +O(\varepsilon^2), \\
		M_1 &=& M_1^{(0)\pm} +\varepsilon M_1^{(1)\pm} +O(\varepsilon^2), \\
			M_3 &=& M_3^{(0)\pm} +\varepsilon M_3^{(1)\pm} +O(\varepsilon^2), \\
	\end{split}
	\end{eqnarray}
where
\begin{eqnarray}
	 M_1^{(0)\pm} &=& \frac{b_H (\tau^{(0)\pm})^2 [(\epsilon^{(0)\pm})^2-3 (\tau^{(0)\pm})^2]}{2(\epsilon^{(0)\pm})^3},  \\
	 M_3^{(0)\pm} &=& \frac{2 b_H (\tau^{(0)\pm})^3}{(\epsilon^{(0)\pm})^2}.  
\end{eqnarray}

The above expanded forms are inserted into Eqs.~\eqref{eq:ep}, \eqref{eq:blm} and \eqref{eq:bam} and like powers of $\varepsilon$ are collected. This provides the leading order equations at O(1),
\begin{eqnarray}
	\begin{split}
		b_\epsilon^{(0)\pm} \epsilon^{(0)\pm} - b_H \frac{(\tau^{(0)\pm})^4}{(\epsilon^{(0)\pm})^3} &=& 0, \\
		\partial_s N_1^{(0)\pm} - \tau^{(0)\pm}  N_2^{(0)\pm} &=& 0, \\
		\partial_s N_2^{(0)\pm} + \tau^{(0)\pm}  N_1^{(0)\pm} &=& 0, \\
		\partial_s N_3^{(0)\pm} &=& 0, \\
	    \partial_s M_1^{(0)\pm}-\tau^{(0)\pm}M_2^{(0)\pm}-N_2^{(0)\pm} &=& 0, \\
		\partial_s M_2^{(0)\pm} +N_1^{(0)\pm}+\tau^{(0)\pm} M_1^{(0)\pm} &=& 0, \\
		\partial_s M_3^{(0)\pm} &=&0, 
	\end{split}
\end{eqnarray}
with the conditions $\tau^{(0)\pm}(0)=\tau_0$ and continuity of all other functions over the inflection point. The last equation gives $M_3^{(0)\pm}$ is constant, and so $(\tau^{(0)\pm})^3 = A^{\pm} (\epsilon^{(0)\pm})^2$, where $A^{\pm}$ is an unknown constant. The first equation then requires $\tau^{(0)\pm}$ to be a constant for $b_\epsilon^{(0)\pm}$ to be constant.  Hence, $\tau^{(0)\pm}=\tau_0$ and $b_\epsilon^{(0)\pm} = (A^{\pm})^2 b_H / \tau_0^2$. Inserting these results into the remaining equations then shows that $N_1^{(0)\pm}=N_2^{(0)\pm}=M_2^{(0)\pm}=0$,  $A^{\pm} =\tau_0/3$, and  $N_3^{(0)\pm}(s)=n_3^{\pm}$, where $n_3^{\pm}$ are constants. In this leading order solution, $\tau$, the first and second components of the force, and all the components of the moment, are continuous over the inflection point. We can also make the third component of the force continuous if we set the constants such that  $n_3^{+}=n_3^{-}=n_3^{(0)}$. The value of $n_3^{(0)}$ depends on the behavior elsewhere along the ribbon. Hence at leading order, only $\epsilon$ is discontinuous, with $\epsilon^{(0)\pm} = \sqrt{3} a^{\pm} |\tau_0|$, where $a^+=1$ on the positive branch and $a^-=-1$ on the negative branch; $a^{\pm}$ is therefore an indicator of the discontinuity in $\epsilon$.

With these results, the O($\varepsilon$) equations become
\begin{eqnarray}
	4 b_H \epsilon^{(1)\pm} + 3 b_H \kappa^{(1)\pm} + \sqrt{3} a^\pm \mbox{ sign}(\tau_0) (9 b_\epsilon^{(1)\pm} \tau_0 - 4 b_H \tau^{(1)\pm})&=&0, \label{eq:delta1} \\
	\partial_s N_1^{(1)\pm} - \tau_0  N_2^{(1)\pm} &=& 0, \label{eq:delta21} \\
	\partial_s N_2^{(1)\pm} + \tau_0  N_1^{(1)\pm} - n_3^{(0)} \kappa^{(1)\pm} &=& 0, \label{eq:delta22}\\
	\partial_s N_3^{(1)\pm} &=& 0, \label{eq:delta3} \\
	\partial_s M_1^{(1)\pm}-\tau_0 M_2^{(1)\pm} -N_2^{(1)\pm} &=&0, \label{eq:delta4}\\
	\partial_s M_2^{(1)\pm}+\tau_0 M_1^{(1)\pm}- \kappa_1^{(1)\pm} M_3^{(0)\pm}+N_1^{(1)\pm} &=&0, \label{eq:delta5}\\
    \partial_s M_3^{(0)\pm} &=&0, \label{eq:delta6}
\end{eqnarray}
where
\begin{eqnarray}
	M_1^{(1)\pm}&=&\frac{b_H }{9}(3 \epsilon^{(1)\pm} +4\kappa^{(1)\pm} - 3 \sqrt{3} a^\pm  \mbox{ sign}(\tau_0)  \tau^{(1)\pm} ), \\
	M_3^{(1)\pm}&=& b_H \frac{-4 \sqrt{3} \epsilon^{(1)\pm} - 3 \sqrt{3}\kappa^{(1)\pm} +18 a^\pm \mbox{ sign}(\tau_0) \tau^{(1)\pm}}{9 a^{\pm}}, \\
\end{eqnarray}
with the conditions $\tau^{(1)}(0)=0$ and $\kappa^{(1)}(0)=0$.
Equations~\eqref{eq:delta1}, \eqref{eq:delta3}, and \eqref{eq:delta6} provide the solutions
\begin{eqnarray}
	\epsilon^{(1)\pm} &=& \frac{- 3 b_H \kappa^{(1)\pm} - 9 \sqrt{3} a^\pm \mbox{ sign}(\tau_0)   b_\epsilon^{(1)\pm} \tau_0 }{4 b_H}, \\
	N_3^{(1)\pm} &=& n_3^{(1)}, \\
	\tau^{(1)\pm} &=& 0,
\end{eqnarray}
where $n_3^{(1)}$ is a constant, and we have used $\tau^{(1)\pm}(0)=0$ and continuity of $N_3$ at $s=0$. The remaining equations can  be written as
\begin{equation}
	\frac{d}{ds} \left(\begin{array}{c}
		\kappa^{(1)\pm} \\
		N_1^{(1)\pm} \\
		N_2^{(1)\pm} \\
		M_2^{(1)\pm}
	\end{array} \right) + \mathbf{A} \left(\begin{array}{c}
		\kappa^{(1)\pm} \\
		N_1^{(1)\pm} \\
		N_2^{(1)\pm} \\
		M_2^{(1)\pm}
	\end{array} \right)  = 	\mathbf{b} , \label{eq:1st order}
\end{equation}
where
\begin{eqnarray}
		\mathbf{A} &=&\left( \begin{array}{c c c c}
			0& 0 & -\frac{36}{7 b_H} & - \frac{36 }{7 b_H} \tau_0 \\
			0 & 0& -\tau_0 & 0 \\
			-n_3^{(0)} & \tau_0 & 0 & 0 \\
			-\frac{17 b_H \tau_0}{36} & 1 & 0 &0
		\end{array}\right), \\
		\mathbf{b} &=& \left(\begin{array}{c}
			0 \\
			0 \\
			0\\
			\frac{3 \sqrt{3}}{4} a^\pm b_\epsilon^{(1)\pm} |\tau_0| \tau_0
		\end{array} \right),
\end{eqnarray}
which is a linear system of first order differential equations with constant coefficients. We note that $\mathbf{b}$ contains the only $a^{\pm}$ term in these equations. This means that the discontinuity does not affect $N_1$, $N_2$ and $M_2$ at this order, as the solution to this equation can be written 
\begin{equation}
	\left(\begin{array}{c}
		\kappa^{(1)\pm}(s) \\
		N_1^{(1)\pm}(s) \\
		N_2^{(1)\pm}(s) \\
		M_2^{(1)\pm}(s)
	\end{array} \right)  = \int_0^{s}  e^{\mathbf{A} (t-s)} \mathbf{b} \,dt + e^{-\mathbf{A} s}  \left(\begin{array}{c}
	\kappa^{(1)\pm}(0) \\
	N_1^{(1)\pm}(0) \\
	N_2^{(1)\pm}(0)\\
	M_2^{(1)\pm}(0)
	\end{array} \right),
\end{equation}
where $e^{\mathbf{A}}$ is the matrix exponential of $\mathbf{A}$, and we have expressed the particular solution as the definite integral from 0 to $s$. The contribution from $\mathbf{b}$ at $s=0$ is therefore 0, so $a^{\pm}$ does not contribute at the inflection point. Enforcing $\kappa^{(1)}(0) =0$ and continuity of the other functions at the point of inflection also means that
\begin{equation}
\left(\begin{array}{c}
	\kappa^{(1)\pm}(0) \\
	N_1^{(1)\pm}(0) \\
	N_2^{(1)\pm}(0)\\
	M_2^{(1)\pm}(0)
\end{array} \right) 
= \left(\begin{array}{c}
0 \\
n_1^{(1)} \\
n_2^{(1)}\\
m_2^{(1)}
\end{array} \right),
\end{equation} 
where  $n_1^{(1)}$, $n_2^{(1)}$, and $m_2^{(1)}$ are unknown constants. With this solution, the first and third moments $M_1^{\pm}$ and $M_3^{\pm}$ are
\begin{eqnarray}
	M_1^{\pm} &=& \varepsilon \left(\frac{7}{36} b_H \kappa^{(1)\pm} - \frac{3\sqrt{3}}{4} a^{\pm} b_\epsilon^{(1)\pm} |\tau_0|\right)  + O(\varepsilon^2),\\
	M_3^{\pm} &=& \frac{2}{3} b_H +3  \varepsilon  b_\epsilon^{(1)\pm} \tau_0 + O(\varepsilon^2).
\end{eqnarray} 
Hence, for $M_1^{\pm}$ to be continuous at $s=0$, we require $ b_\epsilon^{(1)\pm}=0$. This means that $\mathbf{b}=\mathbf{0}$ and the solution to $O(\varepsilon^2)$ becomes
\begin{eqnarray}
\epsilon^{\pm} &=& \sqrt{3} a^{\pm} |\tau(s)| -\frac{3}{4}  \kappa(s)+ O(\varepsilon^2),\\
\tau &=& \tau_0 + O(\varepsilon^2), \\
M_1 &=& \frac{7}{36} b_H  \kappa(s)+ O(\varepsilon^2), \\
	M_3 &=& \frac{2}{3} b_H \tau(s) + O(\varepsilon^2),\\
	N_3 &=& n_3^{(0)} + \varepsilon n_3^{(1)} +O(\varepsilon^2),\\
	b_\epsilon &=& =\frac{b_H}{9} + O(\varepsilon^2),
\end{eqnarray}
and 
\begin{equation}
	\left(\begin{array}{c}
		\kappa(s) \\
		N_1(s) \\
		N_2(s) \\
		M_2(s)
	\end{array} \right)  = e^{-\mathbf{A} s}  \left(\begin{array}{c}
	0 \\
\varepsilon	n_1^{(1)} \\
\varepsilon	n_2^{(1)}\\
\varepsilon	m_2^{(1)}
	\end{array} \right) + O(\varepsilon^2),
\end{equation}
where we have dropped the $\pm$ on the terms that are smooth over the inflection point at $s=0$. The remaining unknown constants, $\varepsilon n_1^{(1)}$, $\varepsilon n_2^{(1)},$ $\varepsilon m_2^{(1)}$, $n_3 =n_3^{(0)} + \varepsilon n_3^{(1)}$, depend on the behavior of the system away from the inflection point. 

\begin{figure}[ht]
	\centering
	\includegraphics[width=0.33\linewidth]{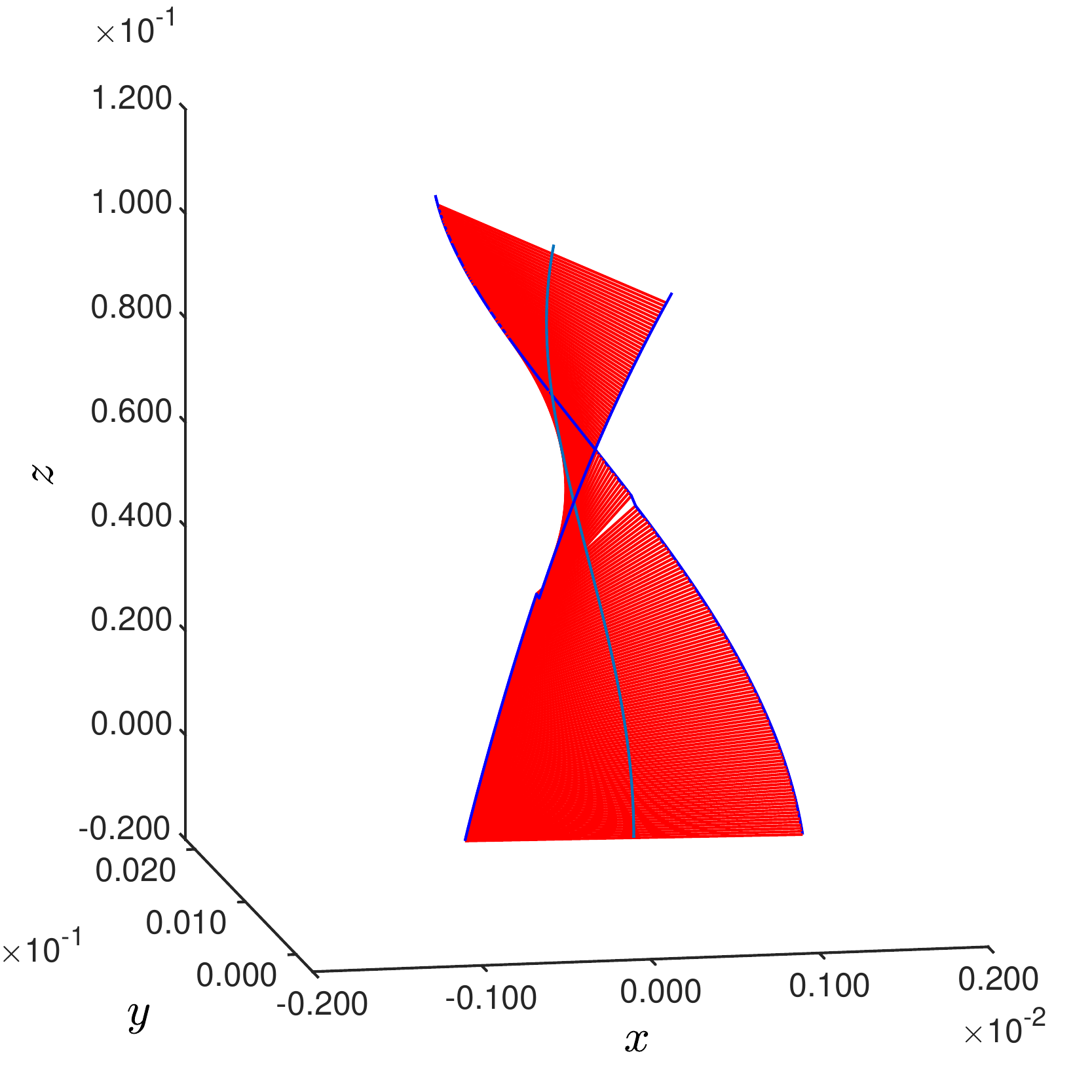} \quad
	\includegraphics[width=0.33\linewidth]{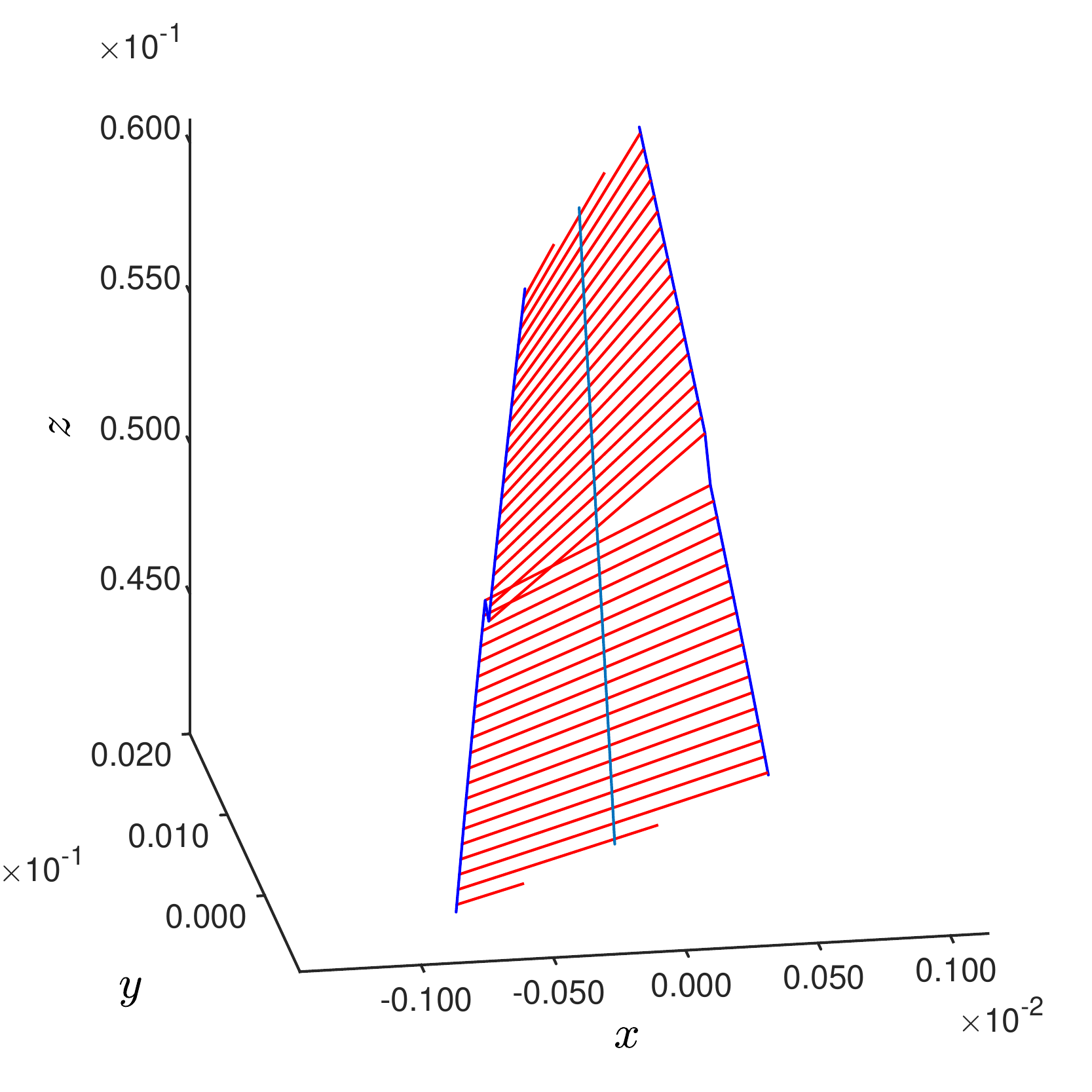} 
	\caption{Inflection point neighborhood in a ribbon of width $w = 2\times10^{-3}$, using $\tau_0 = 20$, $\varepsilon n_1^{(1)}=0$, $\varepsilon n_2^{(1)}= 0.1$, $\varepsilon m_2^{(1)}=0.1$, and an arc-length step size of $\Delta s = 5\times 10^{-4}$. The right image is zoomed in at the inflection point, where $\kappa$ crosses zero. There is a jump in the embedding at this point, as generators change their orientation.} As the ribbon is very thin, we used an aspect ratio different than one to better visualize the generators.
	\label{fig:inflection}    
\end{figure}

Based on this solution, we plot in Fig.~\ref{fig:inflection} the behavior of the ribbon generators at the inflection point, for $b_\epsilon/b_H = 1/9$, a width $w = 2\times10^{-3}$, $\tau_0 = 20$, $n_3^{(0)}=0$, $\varepsilon n_2^{(1)}=0$, $\varepsilon n_2^{(1)}=0.1$, and $\varepsilon m_2^{(1)}=0.1$. 
The jump in the embedding \eqref{eq:jump} $\sim w(b_\epsilon/b_H)^{1/4}$ is visible. We note that this change in ruling orientation due to $f \rightarrow -f$ at $\kappa = 0$ has been explored as an actual constraint at the boundaries when computing solutions to the M\"obius band~\cite{moore2019computation,charrondiere2024merci}.

\section{Conclusion}

A ruled narrow strip model with an auxiliary field provides a potential form of regularization of inflection point singularities, by relaxing the constraint of developability. 
The model provides an energy scaling like that of Freddi and co-workers~\cite{freddi2016corrected}, \com{bridging the the energies of a Sadowsky ribbon and a Kirchhoff rod. For a certain choice of parameters, it is convex and has continuous moment at inflection points, providing} a corrected, convexified Sadowsky functional without any need for patching, with unique small curvature solutions. \com{The energy density prefactors can match those of Freddi and co-workers under a different choice of parameters, but this choice leads to a jump in the moment at inflection points}.
The auxiliary field is zero in regions of large curvature, is nearly linear in curvature in regions of prevailing twist, \com{and suffers jumps at inflection points corresponding to a discontinuity in the embedding proportional to the small width of the ribbon. } 
The equilibrium equations for this model were obtained here using a technique with potential application to other rod models.

\section*{Acknowledgments}
JH thanks Marcelo Dias and Tian Yu for extensive discussions and shared calculations on an earlier approach to this problem. EV and JH thank Abhinav Dehadrai for discussions. 
EV and JH were partially supported by US National Science Foundation grant CMMI-2001262.

\appendix

\section{Euler-Lagrange equations for a Kirchhoff rod}\label{variational}

In Section~\ref{sec:euler}, we derived the Euler-Lagrange equations for the ribbon model using a variational approach from the shell literature~\cite{wisniewski1998shell}.
To further illustrate this approach, we apply it here to the standard model of an inextensible Kirchhoff rod.

Using~\eqref{eq:nd3}, the constrained energy and its variation are
\begin{align}
  \mathcal{E}_{Rod} = \int ds \bigg[\frac{c_1}{2}\kappa_1^2+\frac{c_2}{2}\kappa_2^2+\frac{c_3}{2}\tau^2+\mathbf{N}\cdot(\mathbf{x}'-\mathbf{d}_3)\bigg]\,,
  \\[3mm]
  \delta\mathcal{E}_{Rod} = \int ds \bigg(c_1\kappa_1\delta\kappa_1+c_2\kappa_2\delta\kappa_2+c_3\tau\delta\tau+\mathbf{N}\cdot\delta\mathbf{x}'-\mathbf{d}_3\times\mathbf{N}\cdot\delta\boldsymbol\theta\bigg)\,.
\end{align}

The kinematics of the rod are given by
\begin{align}
  \begin{split}
  \mathbf{d}_1' &= \tau\mathbf{d}_2 -\kappa_2\mathbf{d}_3\,,
  \\[3mm]
  \mathbf{d}_2' &= \kappa_1\mathbf{d}_3-\tau\mathbf{d}_1 \,,
  \\[3mm]
  \mathbf{d}_3' &= \kappa_2\mathbf{d}_1-\kappa_1\mathbf{d}_2 \,.
  \end{split}
\end{align}

Recall that $\bd_i = \mathbf{Q}\cdot\mathbf{D}_i$ and $\delta\bQ\cdot\bQ^\top\cdot(\,) = \delta\boldsymbol\theta\times (\,)$. We have the following variations for the curvatures
\begin{align}
  \begin{split}
    \delta\kappa_1 &= (\delta\bQ\cdot\bD_2)'\cdot\bd_3+\bd_2'\cdot\delta\bQ\cdot\bD_3
    = (\delta\boldsymbol\theta\times\bd_2)'\cdot\bd_3+\bd_2'\cdot\delta\boldsymbol\theta\times\bd_3\,,
    \\[3mm]
    \delta\kappa_2 & = (\delta\bQ\cdot\bD_3)'\cdot\bd_1+\bd_3'\cdot\delta\bQ\cdot\bD_1
    = (\delta\boldsymbol\theta\times\bd_3)'\cdot\bd_1+\bd_3'\cdot\delta\boldsymbol\theta\times\bd_1\,,
    \\[3mm]
    \delta\tau &= (\delta\bQ\cdot\bD_1)'\cdot\bd_2+\bd_1'\cdot\delta\bQ\cdot\bD_2
    = (\delta\boldsymbol\theta\times\bd_1)'\cdot\bd_2+\bd_1'\cdot\delta\boldsymbol\theta\times\bd_2\,.
  \end{split}
\end{align}
We can now substitute these relations into the variation of the rod energy
\begin{align}
\begin{split}
  \delta\mathcal{E}_{Rod} = \int ds \bigg\{&
  c_1\kappa_1\Big[(\delta\boldsymbol\theta\times\bd_2)'\cdot\bd_3+\bd_2'\cdot\delta\boldsymbol\theta\times\bd_3\Big]
+c_2\kappa_2\Big[(\delta\boldsymbol\theta\times\bd_3)'\cdot\bd_1+\bd_3'\cdot\delta\boldsymbol\theta\times\bd_1\Big]
  \\
&+c_3\tau\Big[(\delta\boldsymbol\theta\times\bd_1)'\cdot\bd_2+\bd_1'\cdot\delta\boldsymbol\theta\times\bd_2\Big]
  +\mathbf{N}\cdot\delta\mathbf{x}'-\mathbf{d}_3\times\mathbf{N}\cdot\delta\boldsymbol\theta
  \bigg\}\,, 
\end{split}
\end{align}
and obtain through integration by parts 
\begin{align}
  \begin{split}
  \delta\mathcal{E}_{Rod} =
  \int ds \bigg\{ &
  \bigg[c_1\Big[\kappa_1\bd_3\times\bd_2'-\bd_2\times(\kappa_1\bd_3)'\Big]
  +c_2\Big[\kappa_2\bd_1\times\bd_3'-\bd_3\times(\kappa_2\bd_1)'\Big]
  \\
  &+c_3\Big[\tau\bd_2\times\bd_1'-\bd_1\times(\tau\bd_2)'\Big]-\mathbf{d}_3\times\mathbf{N}\bigg]\cdot\delta\boldsymbol\theta
  -\mathbf{N}'\cdot\delta\mathbf{x}\bigg\}
  \\&+\bigg[(c_1\kappa_1\bd_1+c_2\kappa_2\bd_2+c_3\tau\bd_3)\cdot\delta\boldsymbol\theta+\mathbf{N}\cdot\delta\mathbf{x}\bigg]\bigg|^L_0 \,.
\end{split}
\end{align}
By identifying the boundary term contracted with $\delta\boldsymbol\theta$ as the moment $\mathbf{M}$, we are able to rewrite this variation as
\begin{align}
  \delta\mathcal{E}_{Rod} =
(\mathbf{M}\cdot\delta\boldsymbol\theta+\mathbf{N}\cdot\delta\mathbf{x})\big|^L_0-\int ds\, \bigg[(\mathbf{M}'+\mathbf{d}_3\times\mathbf{N})\cdot\delta\boldsymbol\theta+\mathbf{N}'\cdot\delta\mathbf{x}\,\bigg].
\end{align}

\bibliographystyle{unsrt}
\bibliography{refs_ribbon}

\end{document}